\documentclass[prl,reprint,superscriptaddress,aps]{revtex4-1}

\usepackage[utf8]{inputenc}
\usepackage{graphicx,color}

\definecolor{dark-blue}{rgb}{0,0.2,0.6}

\usepackage[colorlinks,%
            urlcolor=dark-blue,%
            citecolor=dark-blue,%
            linkcolor=dark-blue]{hyperref}

\hypersetup{pdftitle={Observation of coherent multi-orbital polarons in a two-dimensional Fermi gas},%
            pdfauthor={N. Darkwah Oppong, L. Riegger, O. Bettermann, M. Höfer, J. Levinsen, M. M. Parish, S. Fölling, and I. Bloch},
            pdfproducer={},%
            pdfcreator={}}

\usepackage{amsmath}
\usepackage{textcomp}

\usepackage{etoolbox}
\makeatletter
\pretocmd{\NAT@open}{\begingroup\color{\@citecolor}}{}{}
\apptocmd{\NAT@close}{\endgroup}{}{}
\makeatother

\graphicspath{{./figures/}}

\newcommand{\ket}[1]{\ensuremath{\left|{#1}\right\rangle}}

\newcommand{\yb}{\ensuremath{{^\text{173}\textrm{Yb}}}}
\newcommand{\epsF}{\ensuremath{\epsilon_F}}
\newcommand{\kapF}{\ensuremath{\kappa_F}}
\newcommand{\lnkfa}{\ensuremath{\ln(\kapF a_\mathrm{2D})}}

\newcommand{\gket}[1]{\ket{g,#1}}
\newcommand{\gmf}[1]{\gket{m_F=#1}}
\newcommand{\gzero}{\gket{0}}
\newcommand{\gup}{\gket{\uparrow}}
\newcommand{\gdown}{\gket{\downarrow}}

\newcommand{\eket}[1]{\ket{e,#1}}
\newcommand{\emf}[1]{\eket{m_F=#1}}
\newcommand{\edown}{\eket{\downarrow}}

\newcommand{\tP}[1]{\ensuremath{{^3\mathrm{P}_{#1}}}}
\newcommand{\tD}[1]{\ensuremath{{^3\mathrm{D}_{#1}}}}

\newcommand{\sS}[1]{\ensuremath{{^1\mathrm{S}_{#1}}}}

\newcommand{\asciimathunit}[1]{\ensuremath{\,\mathrm{#1}}}

\newcommand{\Erec}{E_\mathrm{rec}}

\newcommand{\nm}{\asciimathunit{nm}}

\newcommand{\kHz}{\asciimathunit{kHz}}

\newcommand{\subfigref}[2]{\hyperref[fig:#1]{\ref*{fig:#1}(#2)}}
\newcommand{\cfigref}[2]{\hyperref[fig:#1]{\ref*{fig:#1}#2}}

\begin{document}


\title{Observation of coherent multiorbital polarons in a two-dimensional Fermi gas}

\author{N.~\surname{Darkwah Oppong}}
\email{n.darkwahoppong@lmu.de}
\author{L.~Riegger}
\author{O.~Bettermann}
\author{M.~H{\"o}fer}
\affiliation{Ludwig-Maximilians-Universit{\"a}t, Schellingstra{\ss}e 4, 80799 M{\"u}nchen, Germany}
\affiliation{Max-Planck-Institut f{\"u}r Quantenoptik, Hans-Kopfermann-Stra{\ss}e 1, 85748 Garching, Germany}

\author{J. Levinsen}
\author{M. M. Parish}
\affiliation{School of Physics and Astronomy, Monash University, Victoria 3800, Australia}

\author{I.~Bloch}
\author{S.~F{\"o}lling}
\affiliation{Ludwig-Maximilians-Universit{\"a}t, Schellingstra{\ss}e 4, 80799 M{\"u}nchen, Germany}
\affiliation{Max-Planck-Institut f{\"u}r Quantenoptik, Hans-Kopfermann-Stra{\ss}e 1, 85748 Garching, Germany}

\date{\today}

\begin{abstract}
  We report on the experimental observation of multiorbital polarons in a two-dimensional Fermi gas of ${}^{173}\mathrm{Yb}$ atoms formed by mobile impurities in the metastable ${^3\mathrm{P}_0}$ orbital and a Fermi sea in the ground-state ${^1\mathrm{S}_0}$ orbital.
We spectroscopically probe the energies of attractive and repulsive polarons close to an orbital Feshbach resonance and characterize their coherence by measuring the quasiparticle residue.
For all probed interaction parameters, the repulsive polaron is a long-lived quasiparticle with a decay rate more than 2 orders of magnitude below its energy.
We formulate a many-body theory, which accurately treats the interorbital interactions in two dimensions and agrees well with the experimental results.
Our work paves the way for the investigation of many-body physics in multiorbital ultracold Fermi gases.
\end{abstract}

\maketitle


The problem of an impurity coupled to a bath lies at the heart of numerous quantum many-body phenomena.
Remarkably, a single localized impurity can dramatically modify the many-body behavior of the medium,
giving rise to intriguing phenomena such as the Kondo effect~\cite{kondo:1964} and Anderson's orthogonality catastrophe~\cite{anderson:1967a}.
Conversely, a mobile impurity interacting with a Fermi sea can form a quasiparticle, or Fermi polaron, with strongly modified properties compared to the bare particle~\cite{massignan:2014}.
The existence of such long-lived quasiparticle states forms the basis of Landau's Fermi liquid theory~\cite{baym:1991}, a paradigm in condensed matter physics.
Moreover, the nature of Fermi polarons has consequences for the stability of itinerant ferromagnetism~\cite{jo:2009,cui:2010,massignan:2011,schmidt:2011,sanner:2012,valtolina:2017,amico:2018}, as well as the phase diagram of spin-imbalanced Fermi gases~\cite{chevy:2006,radzihovsky:2010}.

Ultracold atoms provide an ideal platform for investigating Fermi polarons since impurity-bath interactions can be precisely tuned in the vicinity of a Feshbach resonance, independently of other parameters.
Previously, work in this field was limited to alkali atoms~\cite{nascimbene:2009,schirotzek:2009,kohstall:2012,koschorreck:2012,ong:2015,scazza:2017},
whereas the richer interactions in alkaline-earth(-like) atoms (AEAs) have not yet been utilized.
In AEAs, interactions between the \sS0 ground state (denoted \ket{g}) and the \tP0 ``clock'' state (denoted \ket{e}), a long-lived metastable excited state, are of particular interest since the decoupling of electronic and nuclear degrees of freedom induces spin exchange as well as $\text{SU}(N)$-symmetric collisions~\cite{gorshkov:2010,zhang:2014,scazza:2014}.
The recently discovered orbital Feshbach resonance (OFR) in \yb{}~\cite{zhang:2015,hoefer:2015,pagano:2015} between \ket{g} and \ket{e} has now made it possible to study strongly interacting multiorbital Fermi gases and polarons~\cite{chen:2016,deng:2018,xu:2018,chen:2018}.
Such systems can potentially be used to benchmark theoretical descriptions of interacting fermions in multiple orbitals or bands.
In particular, the multiorbital structure of the OFR means that the closed interaction channel can be strongly affected by the background medium, in contrast to typical Feshbach resonances in alkali atoms.
Thus, an additional Fermi sea in the closed channel is believed to significantly alter the quasiparticle properties~\cite{chen:2018}.

\begin{figure}[b!]
  \includegraphics[width=\columnwidth]{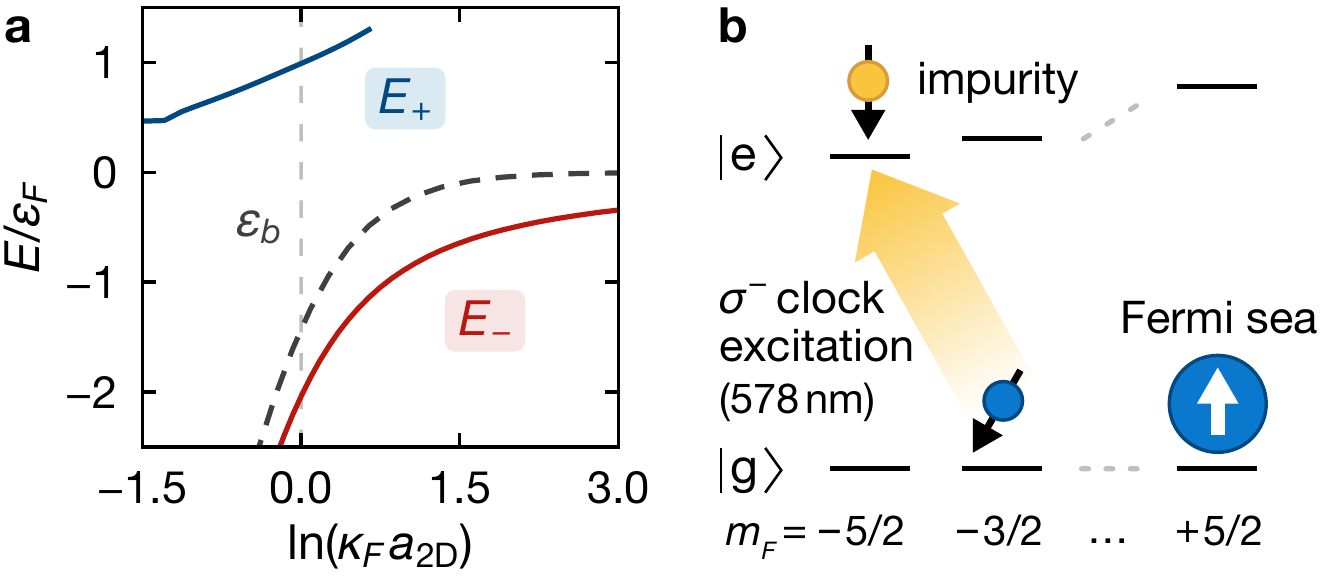}
  \caption{\label{fig:schematic}%
    (a) Numerical calculation of the repulsive polaron (blue line), attractive polaron (red line), and bound dimer (dashed gray line) energies in quasi-2D across the OFR.
    (b) Relevant nuclear spin states in our measurement: The background Fermi sea is in the $m_F=+5/2$  state (blue, \gup{}), while impurities are in the weakly interacting $m_F=-3/2$ ground state (blue, \gzero{}) or the strongly interacting $m_F=-5/2$ ``clock'' state (yellow, \edown{}).
  }
\end{figure}

Fermi polarons in two dimensions are of particular interest due to the increased relevance of quantum fluctuations \cite{zoellner:2011,parish:2011,klawunn:2011}.
Such 2D polarons are also relevant for understanding solid-state systems, as evidenced by the recent observation of Fermi polaron-polaritons in 2D semiconductors~\cite{sidler:2016}.
Similar to three dimensions, there exist attractive and repulsive polaronic branches in 2D~\cite{schmidt:2012,ngampruetikorn:2012} [see Fig.~\subfigref{schematic}{a}].
Despite the successful realization of Fermi polarons in 2D with ultracold gases~\cite{koschorreck:2012,ong:2015}, the coherent nature of the quasiparticles remains largely unexplored, and it is unclear whether the repulsive polaron is well defined given the discrepancy between the theoretically predicted and the experimentally observed polaron energy~\cite{koschorreck:2012}.

Here, we study the many-body physics of Fermi polarons in 2D across the OFR of \yb{}.
We record the spectrum of the many-body system by driving a small number of atoms from a weakly interacting initial state into the strongly interacting final state using the optical clock transition.
We observe two distinct energy branches, which we identify as repulsive and attractive polarons.
By driving Rabi oscillations, we quantify the polaron coherence properties with the quasiparticle residue.
We also investigate the stability of the repulsive polaron against decay into lower lying states, and find that its decay rate remains significantly smaller than the polaron energy even when the latter is a sizable fraction of the Fermi energy.
This relatively large energy and the small decay rate are in contrast to results in 2D with alkali atoms~\cite{koschorreck:2012} and can provide favorable parameters for the stability of ferromagnetism~\cite{massignan:2014}.
We develop a many-body theory for the Fermi polaron in our two-orbital system and find good agreement between theoretical predictions and the experimental results.

In our experiment, we prepare a spin-imbalanced, weakly interacting Fermi gas mixture in the nuclear spin states \gmf{-3/2} (minority, denoted \gzero{}) and \gmf{+5/2} (majority, denoted \gup{}).
After evaporative cooling, we adiabatically ramp up a single-axis optical lattice to a depth of $86\Erec$, where $\Erec = h\times2.0\kHz$ is the lattice recoil energy.
We operate the lattice close to the magic wavelength at $759.35\nm$ ensuring the same trapping potential for \ket{g} and \ket{e} atoms.
The optical lattice generates an array of isolated pancake-shaped traps, where the axial trapping frequency $\omega = 2\pi\times 37.1\kHz$ is much larger than the Fermi energy and kinematics are constrained to 2D.
However, our system is quasi-two-dimensional since the range of the interatomic scattering potential is still smaller than the confinement length scale~\cite{levinsen:2015}.
The external confinement leads to a varying atomic density throughout the trap. We reduce the effects of this inhomogeneity by only considering a small central region, which
contains a few lattice layers and is characterized by an effective background Fermi energy $\epsilon_F \simeq h \times 3.5\kHz$, temperature $T \simeq 0.16 \epsilon_F/k_B$, and minority fraction $N_0 / (N_0 + N_\uparrow) \simeq 0.28$.
The interaction strength between the background Fermi sea in \gup{} and impurities in \emf{-5/2} (denoted \edown) is tuned using a magnetic field in the vicinity of the OFR and is parametrized by \lnkfa{}.
Here, $\kappa_F = \sqrt{2 m \epsilon_F}/\hbar$ is the effective Fermi wave vector, with the \yb{} mass $m$ and the low-energy 2D scattering length $a_\mathrm{2D}$~\cite{SM}%
\nocite{reinaudi:2007,hung:2014,boyd:2007,sansonetti:2005,kitagawa:2008,engelbrecht:1992,kittel:2004,levinsen:2012,parish:2016,chin:2010}.

In the following, we compare our experimental results with a finite-temperature theory of the polaron~\cite{SM} that takes into account the full complexity of the strongly energy-dependent scattering close to the orbital Feshbach resonance~\cite{zhang:2015}, as well as the quasi-2D confinement~\cite{petrov:2001,bloch:2008}.
By considering at most single particle-hole excitations of the background Fermi sea~\cite{chevy:2006,combescot:2007}, we obtain an approximate expression for the impurity self-energy, from which all quasiparticle properties may be determined~\cite{fetter:2003}. Our model has no free parameters and uses the scattering lengths from Ref.~\cite{hoefer:2015}.

In a first experiment, we probe the spectral response of the minority atoms.
We drive atoms from the weakly interacting \gzero{} state into the strongly interacting \edown{} state with a rectangular-shaped pulse addressing the $\sigma^-$ clock transition [see Fig.~\subfigref{schematic}{b}].
The duration is chosen to match a $\pi$ pulse in the absence of interactions, with a Fourier-limited resolution $\simeq 0.1\epsF$ much smaller than the observed linewidths.
The clock laser beam propagates perpendicular to the pancake-shaped traps ensuring that no photon momentum is transferred.
After this spectroscopy pulse, the background Fermi sea and any remaining atoms in \gzero{} are removed from the trap with a ``push'' pulse~\cite{SM}.
Subsequently, the \edown{} impurities are detected by repumping on the $\tD1$ line.
For all parameters, less than $15\%$ of the minority atoms are excited, which leads to a total fraction of \edown{} impurities below $0.05$ and our theoretical description neglects effects of finite impurity density.

\begin{figure}[t!]
  \centering
  \includegraphics[width=\columnwidth]{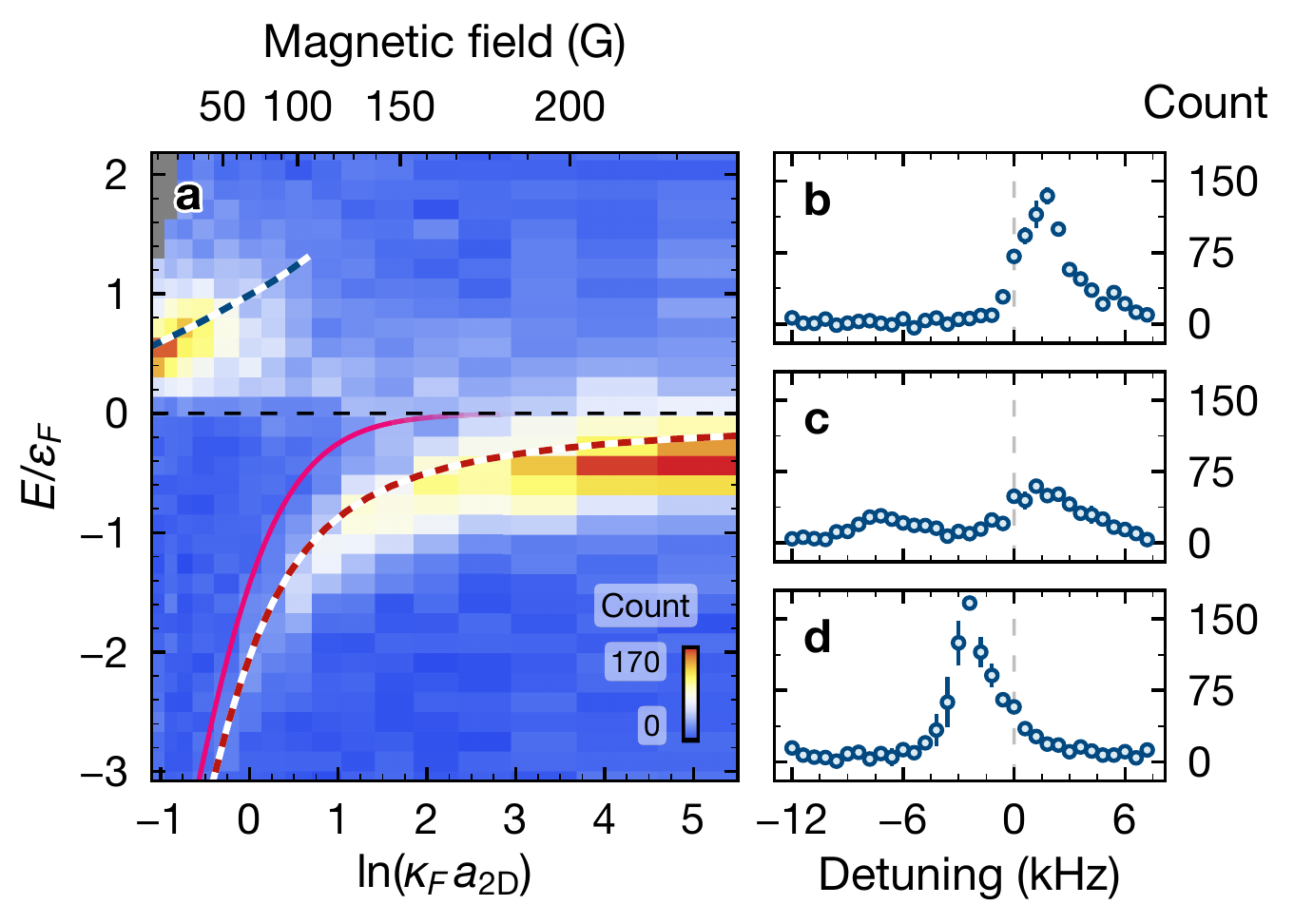}
  \caption{\label{fig:spectrum}%
    Spectral response of the spin-imbalanced Fermi gas across the OFR.
    (a) Number of atoms transferred into \edown{} as a function of detuning $E/\epsilon_F$ and interaction strength \lnkfa.
    Each data point is the average of two or three individual measurements.
    The blue and white (red and white) dashed line denotes the repulsive (attractive) polaron energy predicted from our theory.
    Additionally, we plot the dimer binding energy (solid magenta line) which saturates to $E=0$.
    (b)--(d) Raw spectra recorded with the clock laser at interaction parameters (b) $\lnkfa = -0.72(4)$, (c) $0.26(4)$, and (d) $4.07(6)$.
    Here, zero detuning corresponds to the $\gzero \rightarrow \edown$ transition in the absence of a background Fermi sea.
    Error bars denote the standard error of the mean.
  }
\end{figure}

The measured spectrum is shown in Fig.~\subfigref{spectrum}{a} for $-1.1 \leq \lnkfa \leq 5 $.
Here, we account for the weak repulsion in the initial state [$\lnkfa \simeq -4.9$] by treating the minority atoms in \gzero{} as weakly interacting polarons, and adding the corresponding energy of $0.2\epsF$ to the spectrum~\cite{SM}.
With this calibration, we find a positive energy shift for repulsive interactions [$\lnkfa < 0$] compared to the clock transition without a background Fermi sea.
The energy of this peak increases up to $0.8\epsF$ at $\lnkfa = 0$, beyond which the contrast quickly reduces.
For attractive interactions [$\lnkfa > 0$], we find a second peak at negative energies, with a nonlinear dependence on \lnkfa.
Towards strong interactions [$\lnkfa = 0$], the energy of this peak decreases monotonically to almost $-2.5\epsF$.

The polaron energies, as determined from our theory, are also shown in Fig.~\subfigref{spectrum}{a} and agree very well with the experimental data for both branches.
Therefore, we identify the peak at positive (negative) energy with the repulsive (attractive) polaron.
In Fig.~\subfigref{spectrum}{a}, we also compare the attractive polaron branch to the binding energy of the quasi-2D dimer across the OFR~\cite{SM}.
This energy is significantly higher and we conclude that this state is not addressed for the drive strength used here.
We do not see any direct signatures of a ground-state transition from attractive polaron to bound dimer~\cite{prokofev:2008}, which is predicted for $-1.1 \leq \lnkfa \leq -0.8$ in 2D and for the case of a broad Feshbach resonance~\cite{parish:2011,parish:2013,vlietinck:2014,kroiss:2014}.
However, we do observe a reduction of contrast in the strongly interacting regime until the attractive branch completely vanishes around $\lnkfa \simeq -0.5$.

As illustrated in Figs.~\cfigref{spectrum}{(b)--2(d)}, both branches of the spectrum feature asymmetric line shapes and the peak widths exceed the Fourier limit of the excitation pulse.
The theoretically calculated spectrum approximately reproduces the width of the repulsive branch, but predicts a smaller linewidth for the attractive branch~\cite{SM}.
We believe the excess broadening and asymmetric line shapes are caused by two effects.
First, we average over different background densities due to trap inhomogeneity, which is characterized by the standard deviation of the Fermi energy, $\Delta \epsilon_F \simeq 0.18\epsF$.
Second, the minority fraction prior to excitation is relatively large, with a Fermi energy $\simeq 0.5\epsF$.
This can cause additional broadening and asymmetry due to the width of the minority atoms' initial momentum distribution.

In a second experiment, we extract the quasiparticle residue $Z$, which corresponds to the squared overlap of the polaron and the noninteracting impurity wave function.
The residue quantifies the single-particle coherence of each polaron peak and can be directly determined from the normalized Rabi frequency, $\Omega/\Omega_0 = \sqrt{Z}$~\cite{kohstall:2012}.
Here, $\Omega$ corresponds to the frequency of coherent Rabi oscillations into the interacting state, while $\Omega_0$ is the bare-particle Rabi frequency.
We employ high-intensity clock laser pulses to drive minority atoms into the strongly interacting \edown{} state, yielding an impurity fraction below $0.22$.
The excitation pulse is resonant with the polaron energy and its duration is varied to record the Rabi oscillations.
After the pulse, we detect the remaining minority atoms in the initial \gzero{} state.
To extract the Rabi frequencies, we employ a fit that captures the oscillation frequency $\Omega$ with damping $\Gamma_R$.
The bare-particle Rabi frequency $\Omega_0$ is determined in a similar measurement for each dataset but after removal of the background Fermi sea.
Although large Rabi frequencies, $0.9 < \hbar \Omega_0/\epsF < 1.7$, are required to extract $\Omega$ from a fit, we do not observe a systematic change of $\Omega/\Omega_0$ or $\Gamma_R$ when varying the drive strength $\Omega_0$ in this range~\cite{SM}.

\begin{figure}[t!]
  \includegraphics[width=\columnwidth]{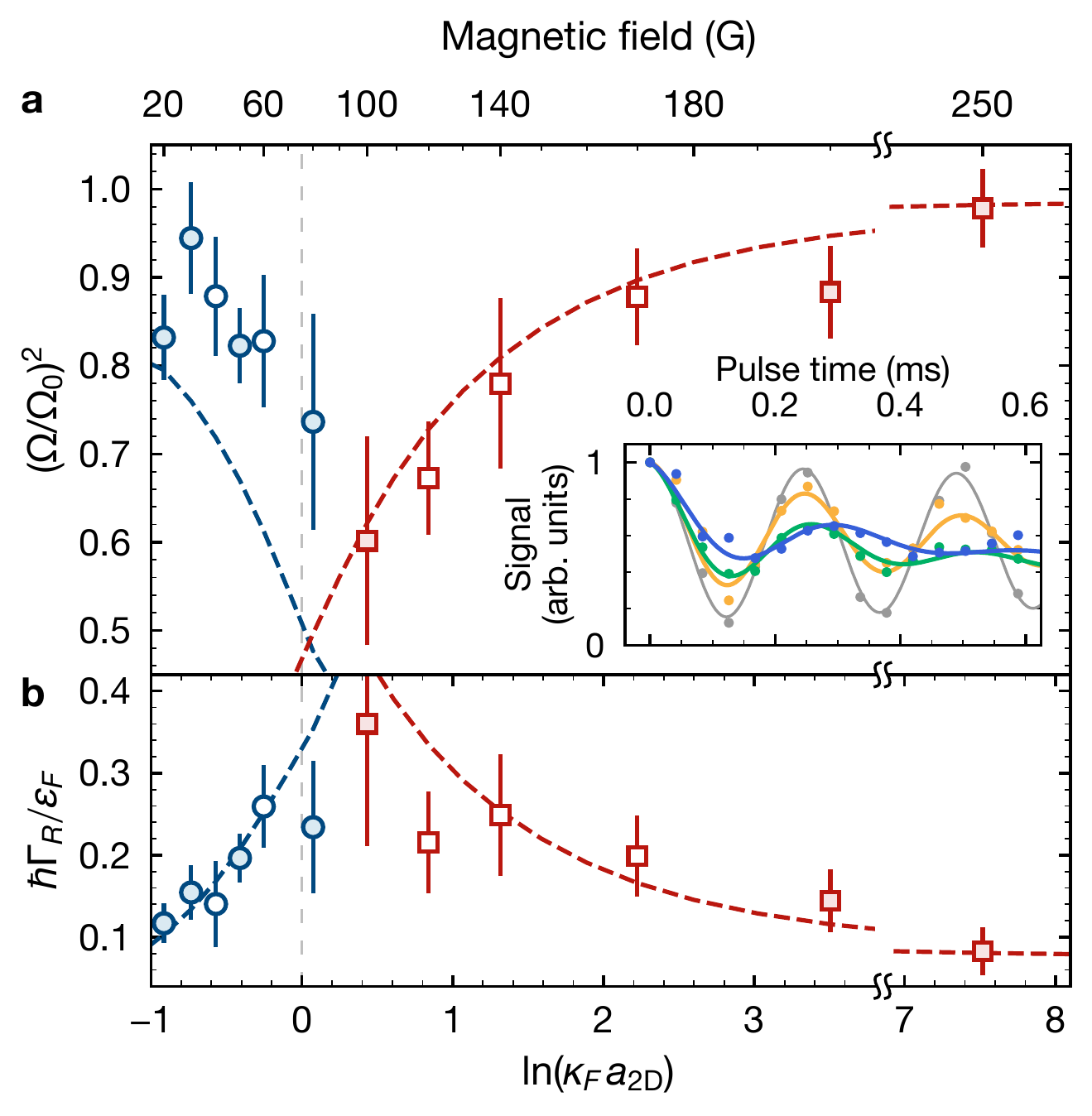}
  \caption{\label{fig:residue}%
    Coherent Rabi oscillations into attractive and repulsive polarons.
    (a) Measured quasiparticle residue $Z = (\Omega/\Omega_0)^2$ extracted from fits to Rabi oscillations.
    Blue circles (red squares) correspond to the repulsive (attractive) polaron.
    The theoretical prediction for $Z$ is shown as dashed lines.
    In the inset, we plot Rabi oscillations at $\lnkfa = -0.41(5)$ (green), $0.84(5)$ (blue), and $7.52(10)$ (yellow) with fits (solid lines).
    The gray points and solid line correspond to a reference measurement without a background Fermi sea.
    (b) Damping rates $\Gamma_R$ of the Rabi coupling into the repulsive (attractive) polaron state are shown as blue circles (red squares).
    The blue dashed line is the width of the repulsive polaron peak from our theory, while the red dashed line corresponds to $(1-Z) + \gamma_0$ with the fitted parameter $\gamma_0 = 0.06(2)$.
    Empty markers refer to data points binned with distinct $\Omega_0$ and error bars denote the fit parameter error.
  }
\end{figure}

In Fig.~\subfigref{residue}{a}, we compare the measured ratio $\left(\Omega/\Omega_0\right)^2$ to the quasiparticle residue $Z$ predicted from theory.
We observe that the normalized Rabi frequencies of both repulsive and attractive polarons decrease towards the strongly interacting regime.
For the attractive polaron, we find very good agreement between our numerical calculation and the experiment.
However, on the repulsive side, the experimentally determined ${\left(\Omega/\Omega_0\right)}^2$ systematically exceeds $Z$, by up to $0.2$.
This discrepancy can be at least partially ascribed to
the finite repulsive interaction in the initial state, which increases the overlap with the repulsive polaron in the final state~\cite{scazza:2017,SM}.

As shown in Fig.~\subfigref{residue}{b}, the damping rate of the Rabi oscillations is large, with $\hbar\Gamma_R \gtrsim 0.1\epsF$, even in the weakly interacting regime.
Remarkably, on the repulsive side, the damping $\hbar \Gamma_R$ matches the width of the repulsive polaron peak in the theoretically calculated spectral function~\cite{SM}, similar to what has been reported in 3D~\cite{scazza:2017}.
This implies that the damping rate of Rabi oscillations is intrinsically connected to the incoherence of the quasiparticle in this case.
By contrast, we expect the attractive polaron to exhibit a smaller degree of incoherence at low temperature since it corresponds to an eigenstate (the ground state at zero temperature)~\cite{yan:2019}.
To model the decoherence of the Rabi oscillations, we use a three-level model featuring the initial state, the attractive polaron, and the continuum represented as a single level with dissipation.
We find that coupling to the continuum leads to a damping proportional to $(1-Z)$~\cite{SM}.
This qualitatively agrees with our observations in Fig.~\subfigref{residue}{b}.

Another important question is the stability of the repulsive polaron with respect to decay into energetically lower lying states such as the attractive polaron or bound dimer.
The polaron decay rate $\Gamma_\mathrm{rep}$ has implications for the realization of itinerant ferromagnetism in the strongly repulsive Fermi gas, since it determines the stability of ferromagnetic domains that may exist when the repulsive polaron energy $E_+ > \epsF$~\cite{massignan:2011,ngampruetikorn:2012,valtolina:2017}.
We measure $\Gamma_\mathrm{rep}$ in the experiment using a double-pulse sequence~\cite{SM} similar to the one successfully applied in Refs.~\cite{kohstall:2012,koschorreck:2012,scazza:2017}:
We use two pulses resonant with the repulsive polaron to drive minority atoms into the strongly interacting \edown{} state, hold them for a variable time, and then drive them back into the initial \gzero{} state for detection.
Immediately after the first pulse, the remaining \gzero{} atoms are removed from the trap before the hold time.
Both the excitation and deexcitation pulse have a bare-particle Rabi frequency $\simeq 0.8\epsF/h$ and the fraction of \edown{} atoms after the deexcitation pulse is between $0.08$ and {$0.16$}.
Since the pulses are spectrally broad and only resonant with the repulsive branch, this measurement captures the decay to lower lying states, while being insensitive to collisional dephasing of the quasiparticles.
We fit the number of remaining impurities to an exponential decay function with a constant offset.

\begin{figure}[t!]
  \includegraphics[width=\columnwidth]{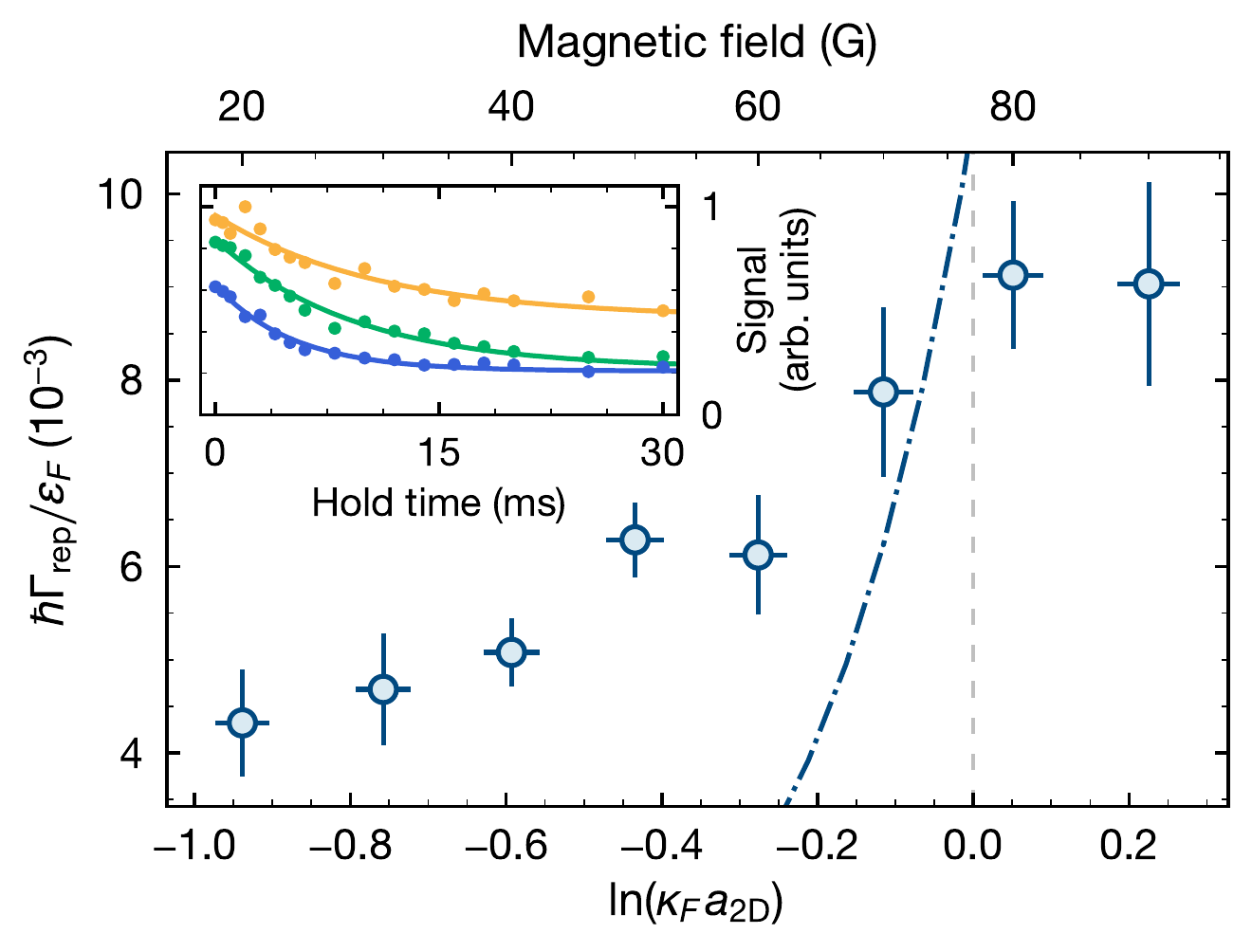}
  \caption{
    Repulsive polaron decay rate (blue circles).
    The error bars denote the fit error in $\Gamma_\mathrm{rep}$ and uncertainty in \lnkfa.
    The blue dash-dotted line corresponds to the decay rate from three-body recombination in 2D for the mean kinetic energy in our system~\cite{SM}.
    The inset shows sample population time traces recorded at $\lnkfa = -0.94(4)$ (yellow), $-0.59(4)$ (green), and $0.05(4)$ (blue), as well as the fitted exponential curves (solid lines).
  }\label{fig:rep_decay}
\end{figure}

In Fig.~\ref{fig:rep_decay}, we display the extracted decay rate $\Gamma_\mathrm{rep}$ as a function of \lnkfa{}.
We observe that it grows monotonically towards strong interactions, but the rate only changes by a factor of 2.
In 2D, a similar weak dependence on the interaction parameter was reported for a broad Feshbach resonance in ${}^\text{40}\text{K}$~\cite{koschorreck:2012}.
However, our observed decay rates are smaller by roughly an order of magnitude with respect to $\epsilon_F/\hbar$.
The inset of Fig.~\ref{fig:rep_decay} shows that the signal settles to a finite plateau, which has also been found in Ref.~\cite{koschorreck:2012}.
In our system, spin exchange at small magnetic fields~\cite{scazza:2014} can potentially populate the \gmf{-5/2} state (denoted \gdown{}), which can contribute to the measured ground-state population.
This could explain the higher plateau for $\lnkfa \simeq -1$, since spin exchange also occurs in the final state after the decay.

We emphasize that the decay rate $\Gamma_\mathrm{rep}$ of the repulsive polaron is several orders of magnitude smaller than the damping rate $\Gamma_R$ for Rabi oscillations in Fig.~\subfigref{residue}{b}.
Therefore, the width of the repulsive polaron peak in the spectral function is set by the dephasing of the repulsive polaron rather than by its decay into the attractive branch.
This is consistent with the findings of Ref. \cite{scazza:2017} but is in contrast to what had been assumed elsewhere, see, e.g., Ref.~\cite{massignan:2014} and references therein.
We approximate the decay process as the recombination of an impurity and two background atoms into a dimer and a free particle.
A similar approximation was successfully used to describe the repulsive polaron decay over several orders of magnitude in 3D~\cite{scazza:2017}.
Here, we consider the three-body recombination rate in a purely 2D geometry~\cite{ngampruetikorn:2013}.
Figure~\ref{fig:rep_decay} shows that the calculated three-body decay agrees with $\Gamma_\mathrm{rep}$ only at $\lnkfa \simeq -0.1$.
However, it strongly disagrees with our experimental result when the dimer becomes larger than the interparticle spacing  [$\lnkfa \gtrsim 0$], or when the dimer is no longer strictly 2D [$\lnkfa \ll 0$].
Notably, $\Gamma_\mathrm{rep}$ in the experiment has a similar functional dependence on \lnkfa{} as the damping rate $\Gamma_R$~\cite{SM}.
This suggests that both the quasi-2D geometry and the medium effects present in a many-body system  need to be included in the calculation of the decay rate, which goes beyond the scope of this work.

The multiorbital nature of interactions in our system introduces the possibility to block intermediate scattering states with the introduction of an additional Fermi sea in \gdown{}.
Our theoretical model predicts increased polaron energies for this configuration, which could potentially stabilize a ferromagnetic phase.
We explore this regime experimentally by preparing a second Fermi sea in \gdown{} with $\epsF = h\times 2.7(2)\kHz$.
In this configuration, we still observe two distinct polaron energy branches, but see only small energy shifts $\leq 0.15\epsF$, within our experimental uncertainties~\cite{SM}.
To address this question, much larger Fermi energies in \gdown{} are desirable, which are currently not accessible in our experiment.

In conclusion, we have realized and comprehensively characterized multiorbital attractive and repulsive Fermi polarons in 2D, which are well described by our many-body theory.
Moreover, we have measured the quasiparticle residue in 2D for the first time.
The particularly long lifetimes of the repulsive polaron in 2D could be beneficial for future studies of itinerant ferromagnetism.
Furthermore, the realization of a multiorbital many-body system with tunable interactions provides a possible platform for the observation of exotic superfluidity, such as the elusive breached-pair phase~\cite{forbes:2005,zou:2018,yu:2019}.
By utilizing tunable mass ratios in state-dependent optical lattices~\cite{riegger:2018}, the present work could be extended to the study of the phase diagram of mass-imbalanced Fermi gases and the regime of Anderson's orthogonality catastrophe~\cite{goold:2011,knap:2012,schmidt:2018}.


\begin{acknowledgments}
  We acknowledge the valuable and helpful discussions with Francesco Scazza and Richard Schmidt, and we thank Vudtiwat Ngampruetikorn for providing his 2D polaron calculation from Ref.~\cite{ngampruetikorn:2012}.
  We also thank Giulio Pasqualetti for building the repump laser system.
  This work was supported by the European Research Council through the synergy grant UQUAM and by the European Union's Horizon 2020 funding.
  N.~D.~O. acknowledges funding from the International Max Planck Research School for Quantum Science and Technology.
  J.~L. and M.~M.~P. acknowledge financial support from the Australian Research Council via Discovery Project No.~DP160102739 and via the ARC Centre of Excellence in Future Low-Energy Electronics Technologies (CE170100039).
  J.~L.~is additionally supported by the Australian Research Council through Future Fellowship FT160100244.
\end{acknowledgments}

\bibliography{references}


\end{document}


\title{{\small Supplemental Material}\\ Observation of coherent multiorbital polarons in a two-dimensional Fermi gas}

\author{N.~\surname{Darkwah Oppong}}
\author{L.~Riegger}
\author{O.~Bettermann}
\author{M.~H{\"o}fer}
\LMU
\MPQ

\author{J. Levinsen}
\author{M. M.~Parish}
\affiliation{School of Physics and Astronomy, Monash University, Victoria 3800, Australia}

\author{I.~Bloch}
\author{S.~F{\"o}lling}
\LMU
\MPQ

\date{\today}

\maketitle

\renewcommand{\thefigure}{S\arabic{figure}}
\setcounter{figure}{0}
\renewcommand{\theequation}{S.\arabic{equation}}
\setcounter{equation}{0}
\renewcommand{\thesection}{S.\Roman{section}}
\setcounter{section}{0}
\renewcommand{\thetable}{S\arabic{table}}
\setcounter{table}{0}

\onecolumngrid

\section{Experimental techniques}

\subsection{State preparation}\label{sec:state-preperation}
\begin{table}[b]
  \begin{ruledtabular}
  \begin{tabular}{l c c c}
    & Fig.~2 & Fig.~3 & Fig.~4 \\\hline
    Fermi energy $\epsilon_F/h$ & $3.65(22)\kHz$ & $3.55(39)\kHz$ & $3.39(21)\kHz$ \\
    Reduced temperature $k_B T/\epsilon_F$ & $0.17(3)$ & $0.16(4)$ & $0.14(3)$ \\
    Minority fraction $C$ & $0.26(2)$ & $0.28(3)$ & $0.31(1)$ \\
    \end{tabular}
  \end{ruledtabular}
  \caption{\label{table:exp-params}%
    Experimental parameters (Fermi energy, temperature, and minority fraction) for each figure presented in the main text.
    The values discussed in the main text are the mean of the corresponding row.%
  }
\end{table}
After forced evaporation in a magic-wavelength ODT (mODT, $\lambda = 759.35\nm$), we load the atoms adiabatically into a magic-wavelength optical lattice with a depth of $86\Erec$, which is ramped up within $500\ms$.
The mODT is kept at a constant value during the optical lattice ramp and holds the atoms against gravity.
Typical trap frequencies in the final trap configuration are $\omega_z \approx 2\pi \times 250\Hz$ along gravity and $\omega_x \approx 2\pi \times 65\Hz$ along the other transverse axis.
Approximating the deep lattice with a harmonic oscillator potential yields an axial trapping frequency $\omega_y = 2\pi \times 37.1\kHz$.
Spin mixtures are prepared by a sequence of intensity-stabilized optical pumping pulses on the intercombination line $\sSF0{5/2} \rightarrow \tPF1{7/2}$ at the beginning of evaporation.
Usually, we prepare an imbalanced $m_F = -3/2, +5/2$ spin mixture in the $\sS0$ ground state (denoted $\gzero$ and $\gup$) with a total atom count $N \approx 50\E3$ and an overall minority fraction $N_{-3/2}/N \approx 0.24$.
Typical temperatures in the mODT before loading into the optical lattice are $T\approx 0.2 E_F/k_B$ determined by fitting density profiles to time-of-flight absorption images.
Here, $k_B$ is the Boltzmann constant.
Ramping up and down the optical lattice leads to an increase of $T$ below $0.05 E_F/k_B$.
The specific experimental parameters for all measurements presented in the main text are shown in Table~\ref{table:exp-params}.

\begin{figure}
  \centering
	\includegraphics[width=\textwidth]{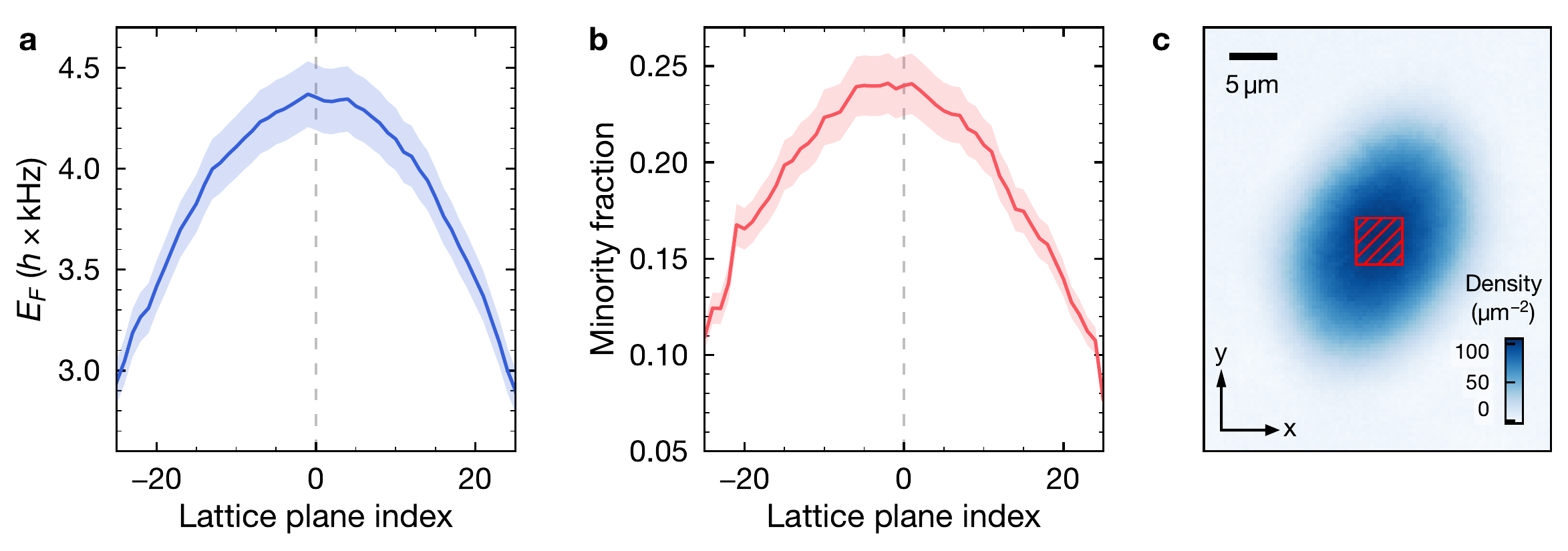}
  \caption{\label{fig:lattice-planes}%
    Characterization of a typical in-trap atomic sample.
    (a) Fermi energy (solid line) and (b) minority fraction (solid line) across different lattice planes.
    The shaded areas denote the error bars of the measured value.
  The image in (c) shows the in-trap density distribution of $m_F=-3/2$ and $m_F=+5/2$ atoms in $\mathrm{atoms}\left/\upmu\mathrm{m}^{2}\right.$.
    The red hatched square denotes a typical integration region and the optical lattice is aligned along the $y$ axis.%
  }
\end{figure}

The optical lattice produces an array of quasi-2D traps for each lattice plane where the confinement $\hbar \omega_y$ is much larger than the 2D Fermi energy $E_F=\hbar \sqrt{2 N \omega_x \omega_z}$.
The harmonic confinement of the optical traps causes each of the lattice planes to be occupied by a different number of atoms.
Fig.~\ref{fig:lattice-planes} shows such a typical distribution extracted from absorption images of the whole atomic cloud.
By imaging the atoms along the transverse direction of the optical lattice, we can select a few lattice planes and reduce the effects of trap inhomogeneity.
The minimum number of selected planes is only limited by the imaging resolution $\approx 3\um$, which is large compared to the lattice spacing ($0.38\um$).
We select a small region with size $\Delta x \times \Delta y = 5.6\um \times 5.6\um $ near the center of the cloud where we integrate the atom number for all our measurements.
This region contains approximately $15$ lattice planes and the variation of atom number per plane is negligible.
Finite temperature as well as the non-zero extent of our integration region lead to an effective Fermi energy $\epsilon_F < E_F$.
We calculate $\epsilon_F$ from the number density $n(x, z)$ in the local density approximation,
\begin{align}
  n(x, z) &= -\frac{m}{2\pi \hbar^2 \beta} \mathrm{Li}_{1}\left[ -e^{-\beta \left( \frac{m}{2}\left[ \omega_x^2 x^2 + \omega_z^2 z^2 \right] - \mu\right)} \right],%
\label{eq:number-density}
\end{align}
where $\mathrm{Li}_{s}(x)$ is the polylogarithm function of order $s$, $m$ is the mass of \yb{}, $\mu$ denotes the chemical potential, and $\beta = 1 / k_B T$.
The effective background Fermi energy $\epsilon_F$ sampled by the minority atoms is then given by
\begin{align}
  \epsilon_F &= \frac{1}{N_0(\Delta x)} \int_{-\Delta x/2}^{\Delta x/2}dx \int_{-\infty}^\infty dz\;n_0(x, z)\, E_F(x, z).
\end{align}
Here, $N_0(\Delta x) = \int_{-\Delta x/2}^{\Delta x/2}dx\int_{-\infty}^\infty dz\;n_0(x, z)$ is the total number of the $\gzero$ atoms with $n_0(x, z)$ the corresponding number density, and
\begin{align}
  E_F(x, z) = \frac{\hbar^2}{2m}\left[ 4\pi n_\uparrow(x, z) \right]
\end{align}
is the local Fermi energy of the background atoms with $n_\uparrow(x, z)$ the number density of the $\gup$ atoms.
We calculate the standard deviation $\Delta \epsilon_F$ to quantify the variation of effective Fermi energy in our integration region,
\begin{align}
  {\left(\Delta \epsilon_F\right)}^2 = \frac{1}{N_0(\Delta x)}\int_{-\Delta x/2}^{\Delta x/2}dx \int_{-\infty}^\infty dz\;n_0(x, z) {\left[ \epsilon_F - E_F(x, z) \right]}^2.
\end{align}
Typically, we find $\epsilon_F \simeq 0.8 E_F$ and $\Delta \epsilon_F \simeq 0.2\epsilon_F$. For the effective Fermi wavevector $\kappa_F$, we use
\begin{align}
  \kappa_F = \frac{\sqrt{2m \epsilon_F}}{\hbar}.
\end{align}
For the minority fraction $C$, we calculate
\begin{align}
  C &= \frac{N_0(\Delta x)}{N_0(\Delta x) + \int_{-\Delta x/2}^{\Delta x/2}dx \int_{-\infty}^\infty dz\;n_\uparrow(x, z)} .
\end{align}

\subsection{In-situ thermometry}
\begin{figure}
  \centering
	\includegraphics[width=0.5\textwidth]{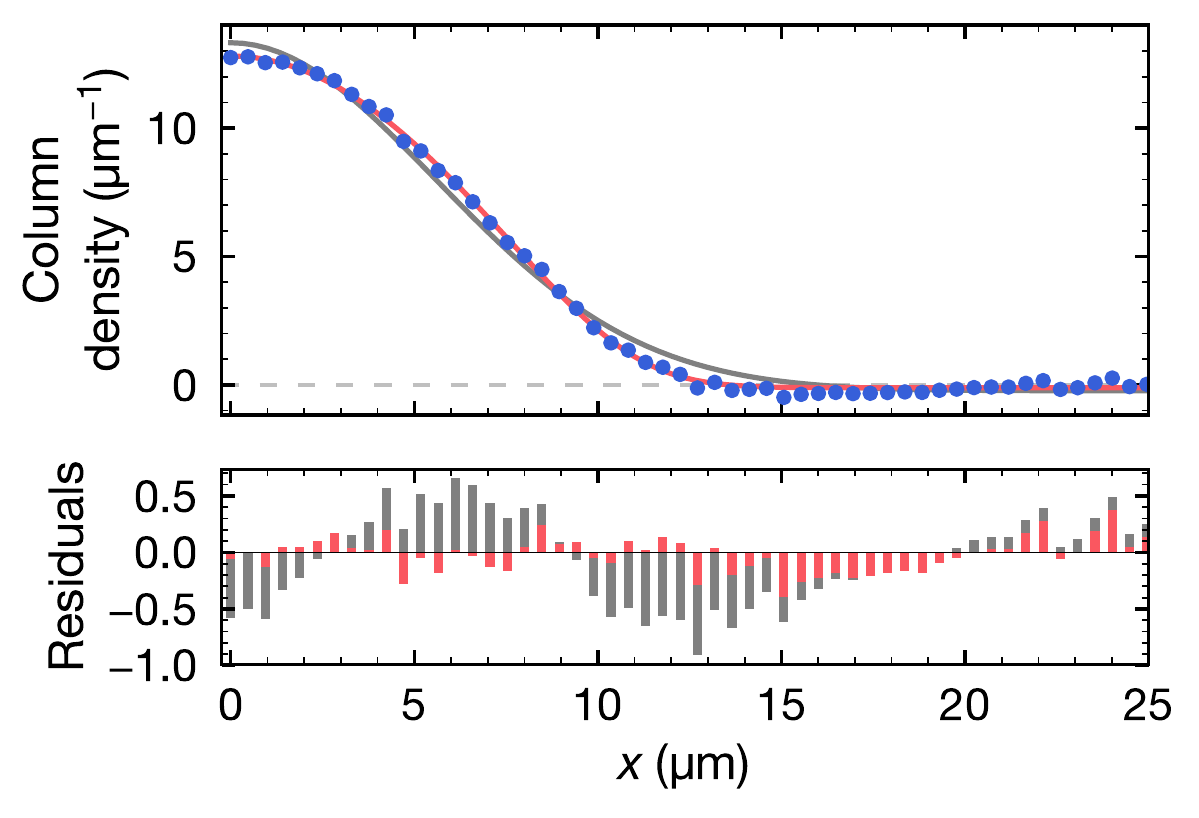}
  \caption{\label{fig:fermi-fit}%
    The upper plot shows fits to the in-situ column density (blue circles) using Eq.~\eqref{eq:column-density} ({\em Fermi-Dirac}, red solid line) and a Gaussian function ({\em Maxwell-Boltzmann}, gray solid line).
    Fit residuals both for the Fermi-Dirac (red bars) and Maxwell-Boltzmann (gray bars) fits are shown in the lower plot.
    The minority fraction in this data set is 0.08(1) and the fit yields $T/T_F = 0.18(2)$.%
  }
\end{figure}
For extracting temperatures in the 2D planes of the optical lattice, we use a fit to the  column density $n(x)$, which we find by integrating Eq.~\eqref{eq:number-density} along $z$,
\begin{align}
  n(x) &=  -\frac{\sqrt{m}}{\sqrt{2\pi} \hbar^2 \beta^{3/2}\omega_y}  \,\mathrm{Li}_{3/2}\left( -e^{-\beta \left[ \frac{m}{2} \omega_x^2 x^2 - \mu\right]} \right).%
  \label{eq:column-density}
\end{align}
Here, the fugacity $z = e^{\beta\mu}$ is related to the temperature by
\begin{align}
  T = \frac{E_F/k_B}{\sqrt{-2 \mathrm{Li}_2 (-z)}}.
\end{align}
In the experiment, we cannot reliably measure the in-situ densities of $\gzero$ and $\gup$ atoms separately.
Instead, we fit the distribution of all atoms.
Simulating density profiles for imbalanced, non-interacting samples shows that fitting the combined density leads to a small overestimation of the temperature.
However, we cannot confirm this result when we compare fitted temperatures at the edge of the cloud (small minority fraction) with the center of the cloud (large minority fraction).
This can possibly be explained with the non-zero ground state interaction which leads to a small broadening of the density profile across the trap.
We perform the temperature fit across multiple planes of the optical lattice and average the results to retrieve the absolute temperature.
The standard deviation is used as the error in the temperature fit.
Fig.~\ref{fig:fermi-fit} shows such a fit at the edge of the cloud where the minority fraction is small.
We relate the temperature to the effective Fermi energy $\epsilon_F$ which yields the reduced temperature $k_B T/\epsilon_F$ (see Table~\ref{table:exp-params}).

\subsection{Atom number calibration}
\begin{figure}
  \centering
	\includegraphics[width=\textwidth]{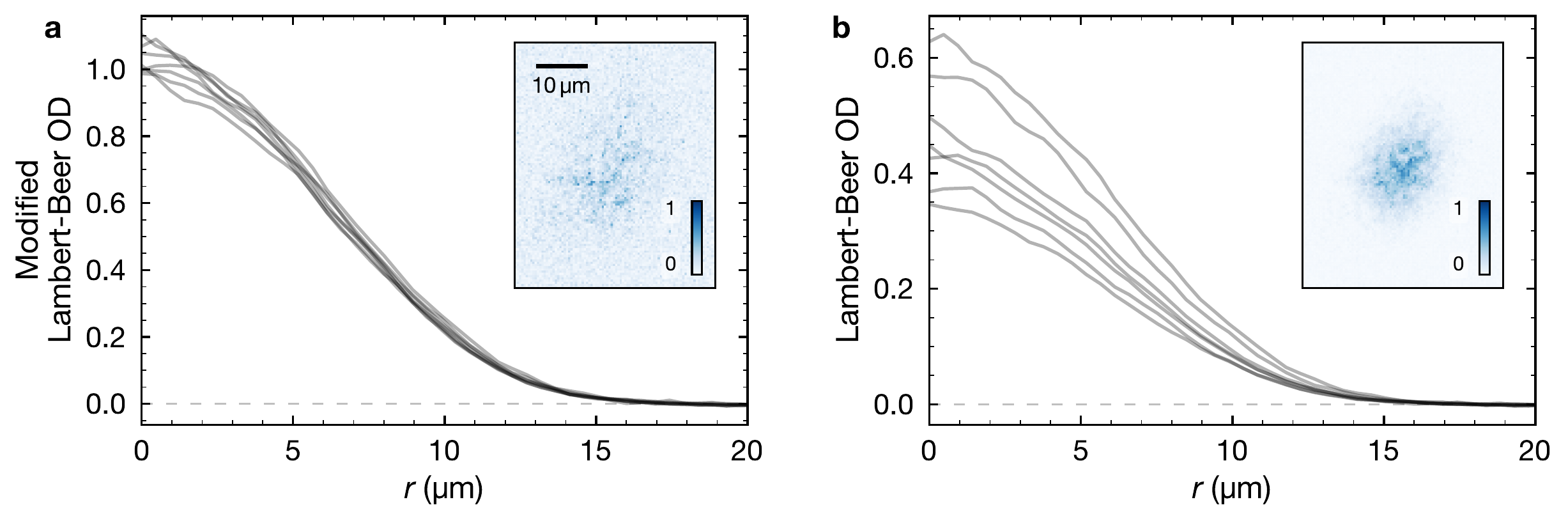}
  \caption{\label{fig:ods-comparison}%
    Azimuthal averages of the optical column density for a spin-polarized, $m_F=-3/2$ sample recorded with different imaging intensities between $2.6 I_\mathrm{sat}$ and $6.2 I_\mathrm{sat}$.
    The data in (a) contains the corrections from the modified Lambert-Beer law [See Eq.~\eqref{eq:modified-lambert-beer}] with $\alpha = 2.673$ whereas (b) does not contain any corrections and shows a strong dependence on the imaging intensity.
    The insets show the variance of the atomic density normalized to the average atom count per pixel.%
  }
\end{figure}
Measurement of the absolute atom number is crucial for determining the Fermi energies in our experiment.
All atom number measurements are done with in-situ imaging at a small magnetic bias field of $1\Gauss$.
We use the broad $\sSF0{5/2} \rightarrow \sPF1{7/2}$ transition and address the $\sigma^+$ as well as $\sigma^-$ transition.
Short pulses with a duration of $15\us$ minimize atomic motion during absorption imaging.
However, the short imaging pulses require intensities larger than the saturation intensity $I_\mathrm{sat}$ to fully penetrate the dense atomic cloud.
We account for the saturated atomic transition with the modified Lambert-Beer law~\cite{reinaudi:2007}
\begin{align}
  \mathrm{OD}(i, j) = \log \left[ \frac{I_\mathrm{in}(i, j)}{I_\mathrm{out}(i, j)} \right] +\frac{I_\mathrm{in}(i, j) - I_\mathrm{out}(i, j)}{ \alpha  I_\mathrm{sat}}.%
  \label{eq:modified-lambert-beer}
\end{align}
Here, $\mathrm{OD}(i, j)$ is the optical column density on camera pixel $(i, j)$ and $I_\mathrm{in}$ is the incident intensity whereas $I_\mathrm{out}$ is the intensity after absorption.
The parameter $\alpha > 1$ accounts for imperfect polarization of the imaging light and the multi-level nature of the atom.
This particularly applies to \yb{}, since imaging on the $^1\mathrm{S}_0^{F=5/2} \rightarrow\; {^1\mathrm{P}_1^{F=7/2}}$ transition involves up to six $m_F$ states.
In the experiment, we determine $\alpha$ by imaging almost identical atomic samples with varying intensity.
Minimization of the atomic density variance across all pixels then yields $\alpha$.
This parameter is independently determined for spin-polarized samples with $m_F \in \{-5/2, -3/2, +5/2\}$ and a weighted average is used for imaging spin mixtures.
Additionally, we measure $\alpha$ for the spin mixture which results from repumping $m_F = -5/2$ in the clock state back to the ground state on the $\tPF0{5/2} \rightarrow \tDF1{7/2}$ transition.
We use the methods of Ref.~\cite{hung:2014} to determine the effective intensity $I_\mathrm{in}(i, j)$ at the location of the atoms and find agreement within $11\%$ of the directly measured intensity.

\subsection{Orbital Feshbach resonance}
We use the description in Ref.~\cite{zhang:2015,hoefer:2015} for the orbital Feshbach resonance.
In this formalism, the Feshbach resonance is fully characterized by the scattering lengths $a_\pm$, the effective ranges $r_\pm$, and the differential Zeeman shift $\Delta \mu$ between $\gup$ and $\edown$. We use the scattering lengths and effective ranges experimentally determined in Ref.~\cite{hoefer:2015}
\begin{align}
  a_+ &= 1878(37)\,a_0,&\; a_- &= 219.7(2.2)\,a_0, \\
  r_+ &= 216\,a_0,&\; r_- &= 126\,a_0,
\end{align}
where $a_0$ is the Bohr radius.
We measure $\Delta\mu$ with the methods from Ref.~\cite{boyd:2007}.
This differential measurement is independent of the magnetic field calibration and only depends on the nuclear magnetic moment of \yb{}, $\mu_I = -0.6776\mu_N$~\cite{sansonetti:2005}.
We find $\Delta \mu\left/ \Delta m_F \right. = h \times 110.72(51)\,\mathrm{Hz}/\mathrm{G}$, where $\Delta m_F$ is the difference between the ground and excited state $m_F$ numbers.
We also use this value for the calibration of the magnetic fields in our experiment.
For the $m_F = \pm 5/2$ states used in all measurements, the differential Zeeman shift is given by
\begin{align}
  \Delta \mu = h \times 554(3)\,\mathrm{Hz/G}.\label{eq:diff-zs-shift}
\end{align}

\subsection{Clock spectroscopy}%
\label{sec:clock-spectroscopy}
\begin{figure}
  \centering
	\includegraphics[width=\textwidth]{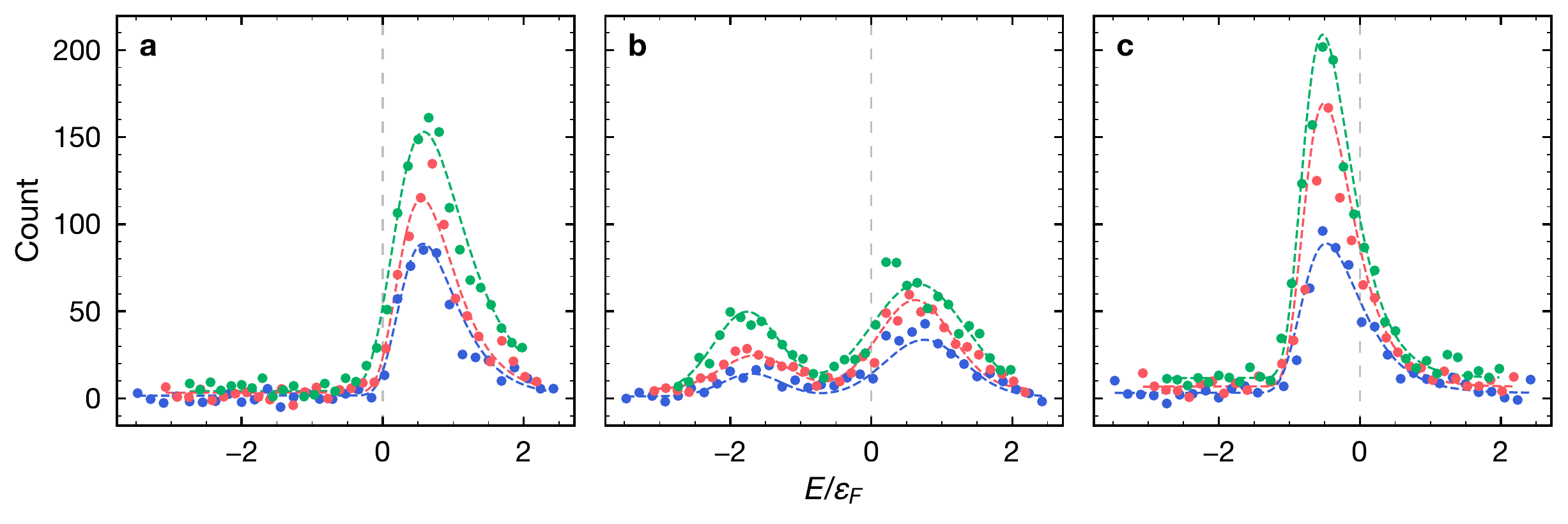}
  \caption{\label{fig:roi-spectrum}%
    Clock spectroscopy at magnetic fields (a) $30\Gauss$, (b) $90\Gauss$, and (c) $210\Gauss$ across different integration regions with varying effective Fermi energy $\epsilon_F = h\times 4.1(2)\kHz$ (green circles), $h\times 3.7(2)\kHz$ (red circles), and $h\times 3.3(2)\kHz$ (blue circles).
    Due to different $\kappa_F$ in each region, the interaction parameter differs by~$\approx 0.1$ and the mean $\ln\left(\kappa_F a_\mathrm{2D}\right)$ is given by (a) $-0.6$, (b) $0.4$, and (c) $4.2$.
    Note that the effective minority fraction varies across the regions as well, with $c = 0.33(3)$ (green circles), $0.28(3)$ (red circles), and $0.24(3)$ (blue circles).
    Dashed lines are Gumbel distribution fits [(a) and (c)] or Gaussian fits (b) intended as a guide to the eye.
    Each data point is the average of two or three individual measurements and error bars are not shown to reduce visual clutter.%
  }%
\end{figure}
We use circularly polarized light to address the $m_F=-3/2 \rightarrow m_F=-5/2$ transition.
After the clock excitation pulse, we rapidly lower the magnetic field to $1\Gauss$ and turn on a high-intensity ($I \sim I_\mathrm{sat}$) ``push'' beam on the broad $\sSF0{5/2} \rightarrow \sPF1{7/2}$ transition.
This removes all ground state atoms from the trap.
Subsequently, the impurity atoms in the clock state are pumped back to the ground state with a $0.3\ms$ pulse on the $\tPF0{5/2} \rightarrow \tDF1{7/2}$ transition and are imaged.
During the repump pulse, atoms can move due to finite photon recoil transferred during the cascade decay $\tD1 \rightarrow \tP1 \rightarrow \sS0$.
We estimate the worst case transverse motion of a single atom during the $0.3\ms$ pulse to be $1.7\um$, which is below our imaging resolution.
Our repumping efficiency is $0.86(4)$ which we account for when counting atoms in the clock state.
We select a small central region for counting atoms as described in Section~\ref{sec:state-preperation}.
When varying the position of the integration region, we can probe planes of the optical lattice with smaller effective Fermi energy $\epsilon_F$ and minority fraction which is shown in Fig.~\ref{fig:roi-spectrum}.
However, we do not find any systematic energy shifts within our experimental resolution for regions with lower minority fractions.
We also do not find any significant modification of the observed lineshapes for regions with different effective Fermi energies.

The initial ground-state preparation in our measurement has weak repulsive interactions which results in a positive energy shift.
We account for this effect by estimating the ground state interaction energy and shifting the spectrum accordingly.
For the calculation of this energy shift, we consider the ground state minority atoms as weakly interacting repulsive polarons with energy $E_+$.
In the limit of a deeply bound dimer ($\eb \gg \epsilon_F$), the energy is given by~\cite{ngampruetikorn:2012}
\begin{align}
	\frac{E_+}{\epsilon_F} \approx \frac{2}{\ln(\eb/\epsilon_F)} = -\frac{1}{\ln(\kappa_F a_{\mathrm{2D},\,gg})}.
\end{align}
Here, $\eb$ is the dimer binding energy and the 2D scattering length of the ground state is given by~\cite{petrov:2001,bloch:2008}
\begin{align}
	a_{\mathrm{2D},\,gg} = l_y \sqrt{\frac{\pi}{D}} \exp\left(-\sqrt{\frac{\pi}{2}}\, \frac{l_y}{a_{gg}}\right) \simeq 12.5 \,a_0,
\end{align}
where $a_{gg} = 199.4a_0$ is the ground state s-wave scattering length~\cite{kitagawa:2008}, $a_0$ is the Bohr radius, $l_y=\hbar/\sqrt{m\omega_y}$ is the oscillator length, and $D \simeq 0.905$.
We find $\ln(\kappa_F a_{\mathrm{2D},\,gg}) = -4.9(1)$ and $E_+ = 0.2\epsilon_F$ for the repulsive polaron in the ground state.
In Fig.~2 of the main text, we use this energy $\Delta E = E_+$ to shift the experimental data in the spectrum.
However, when comparing theory and experiment at large magnetic fields where the attractive polaron is only weakly interacting, we recover shifts which could be explained by a larger ground state interaction, i.e. $0.4(2)\epsilon_F$ at $\ln\left(\kappa_F a_\mathrm{2D}\right)=8.2(1)$.
This disagreement could originate from systematic errors due to the spectral lineshape fit as well as effects of finite momentum and temperature for the polaron energy.

Our clock laser is stabilized to an optical cavity and the residual, non-linear drift of this laser over the course of recording a spectrum such as that shown in Fig.~2 of the main text is typically $\sim 100\Hz$ which is below the relevant energy scales of the experiment.
We can probe single-particle Fourier-limited line shapes down to $200\Hz$ with the configuration discussed in Section~\ref{sec:state-preperation}.
The deep magic-wavelength lattice ensures recoil-free clock spectroscopy.
Residual misalignment between clock laser and optical lattice is below $26\mrad$ and leads to a negligible transverse recoil energy of less than $ h \times 2 \Hz$.

\begin{figure}
  \centering
	\includegraphics[width=\textwidth]{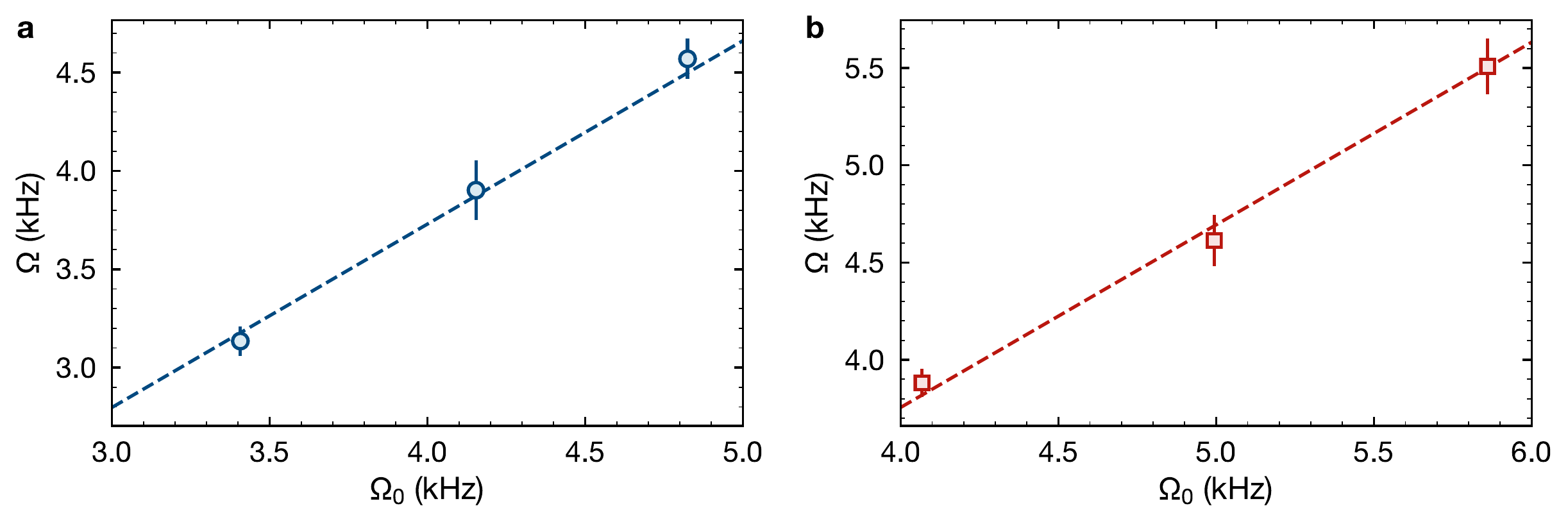}
  \caption{\label{fig:rabi-freq-linearity}%
    Rabi frequencies $\Omega$ for different bare Rabi couplings $\Omega_0$ at (a) $\ln\left(\kappa_Fa_\mathrm{2D}\right) = -0.57(5)$ (repulsive polaron, blue circles), and (b) $\ln\left(\kappa_Fa_\mathrm{2D}\right) = 2.22(6)$ (attractive polaron, red squares).
    The error bars denote the errors of the fit to the Rabi oscillations.
		Dashed lines are linear fits to $\eta \times\Omega_0$.
		The extracted parameter $\eta$ is (a) $0.93(1)$ and (b) $0.94(1)$.%
  }%
\end{figure}

\subsection{Rabi oscillations}\label{sec:rabi-oscillations}
After the clock excitation pulse, we detect the remaining $\gzero$ atoms by first removing the $\gup$ atoms with a resonant pulse on the intercombination line ($250\us$ duration).
The detection of remaining $\gzero$ atoms works more robustly for us than repumping the impurities in $\edown$.
We fit the detected atom number $N$ as a function of the pulse duration $t$ to
\begin{align}
	N(t) = a e^{-\Gamma_{bg} t} - b e^{-\Gamma_R t} \cos(\Omega t).%
  \label{eq:rabi-oscillations-fit-function}
\end{align}
Here, $\Gamma_{bg}$ is the background decay rate accounting for losses in the excited state, $\Gamma_R$ is the coherence decay rate, $\Omega$ is the Rabi frequency, and $a$, $b$ are two dimensionless parameters.
After every experimental cycle, we prepare a non-interacting, spin-polarized sample ($m_F=-3/2$), and measure the single-particle Rabi frequency $\Omega_0$.
For this, we use a fit function without a background decay rate and for the decoherence rate we typically find $\Gamma_R \approx 0.3\kHz$.
The non-zero $\Gamma_R$ can be attributed to finite laser linewidth and other experimental imperfections.
We find $\Omega \propto \Omega_0$ for various single-particle Rabi frequencies $\Omega_0$ both on the repulsive and attractive side as shown in Fig.~\ref{fig:rabi-freq-linearity}.
On the repulsive side at small magnetic fields, the large Rabi frequencies are problematic since the clock transitions of other spin states can be addressed as well.
The detuning of the $m_F=+5/2 \rightarrow {m_{F}}^\prime = +3/2$ transition is $\simeq 0.55 \asciimathunit{kHz/G}$.
We verify that addressing of this additional transition is negligible for magnetic fields $\geq$ $20\Gauss$ by detecting atoms driven to the $m_F=+3/2$ clock state in an independent measurement.

\subsection{Repulsive polaron lifetime}%
\begin{figure}
  \centering
	\includegraphics[width=\textwidth]{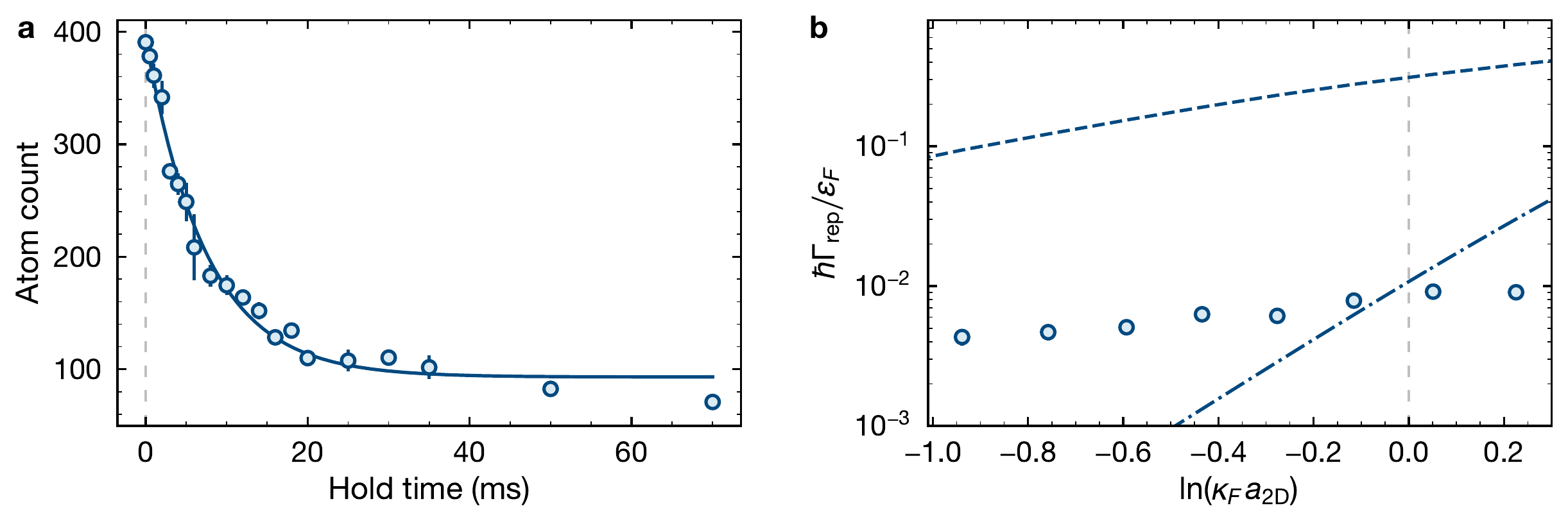}
  \caption{\label{fig:rep-decay}%
    (a) Repulsive polaron decay at $\ln(\kappa_F a_\mathrm{2D}) = -0.43(4)$.
    The blue circles correspond to the number of remaining atoms in the $\gzero$ state after the double-pulse sequence.
    Each data point is the mean of two separate measurements and the error bars refer to standard error of the mean.
    The solid blue line is a fit to an exponential decay function with a constant offset.
    (b) Comparison of experimentally measured $\Gamma_\mathrm{rep}$ (see Fig.~4 of the main text) with theoretical predictions.
    The dash-dotted line is the decay rate due to three-body recombination and the dashed line is the width of the repulsive polaron branch.
    Data points are shown as blue circles and the error bars denote the uncertainty in the fit.%
  }
\end{figure}

\label{sec:repulsive-polaron-lifetime}
We measure the repulsive polaron lifetime with two sequential clock pulses.
Right after the first pulse, we remove any remaining $\gzero$ atoms with a sequence of ``push'' pulses on the intercombination line ($70\us$ duration).
The total number of impurities transferred back to $\gzero$ is detected after the second pulse and after removing all majority atoms ($\gup$) using another ``push'' pulse ($100\us$ duration).
We fit the detected atom number $N$ with
\begin{align}
	N(t) = a e^{-\Gamma_\mathrm{rep} t} + b,
\end{align}
where $\Gamma_\mathrm{rep}$ is the decay rate and $a$, $b$ are dimensionless fit parameters.
Fig.~\subfigref{rep-decay}{a} shows such a fit for longer times $t$ compared to the inset in Fig.~4 of the main text.
Here, the fast initial decay and the slow decay of the plateau is clearly visible.
In Fig.~\subfigref{rep-decay}{b}, we plot the data of Fig.~4 of the main text along with theoretically expected contributions to the repulsive polaron decay rate.
We note that the width of the repulsive polaron branch predicted from our polaron theory model in Eq.~\eqref{eq:width} (see below) approximately matches the slope, but its mean amplitude disagrees with the experimental data.
We also compare with the 2D three-body recombination rate $K_3$ from Ref.~\cite{ngampruetikorn:2013}, where we assume that the background density $n_\uparrow$ is constant such that we obtain the decay rate
through $dn_\downarrow/dt = n_\downarrow [n_\uparrow^2 K_3(E)] = n_\downarrow \Gamma_{\rm rep}$ for the impurity density $n_\downarrow$.
To estimate the collision energy $E$, we assume a zero momentum impurity and approximate $E$ as the mean kinetic energy of the background sampled by the minority atoms,
\begin{align}
  \epsilon_\mathrm{kin} &= \frac{1}{\,N_0(\Delta x)} \int_{-\Delta x/2}^{\Delta x/2}dx \int_{-\infty}^\infty dz\;n_0(x, z)\,E_\mathrm{kin,\uparrow}(x, z),
\end{align}
where the local kinetic energy $E_\mathrm{kin,\uparrow}(x, z)$ is given by
\begin{align}
  E_\mathrm{kin,\uparrow}(x, z) = \frac{2\pi}{n_\uparrow(x, z)} \int_0^\infty dp\, p \frac{p^2}{2m} f_\uparrow(x, z, p).
\end{align}
Here, $f_\uparrow(x, z, p) = \left\{1 + e^{\beta \left[m/2\,\left(\omega_x^2x^2 + \omega_z^2z^2\right) + p^2/(2m) - \mu\right]}\right\}^{-1}$ is the $\gup{}$ number density in phase space.
%

\subsection{Polarons with dual Fermi seas}%
\label{sec:twoFS}
\begin{figure}
  \centering
	\includegraphics[width=\textwidth]{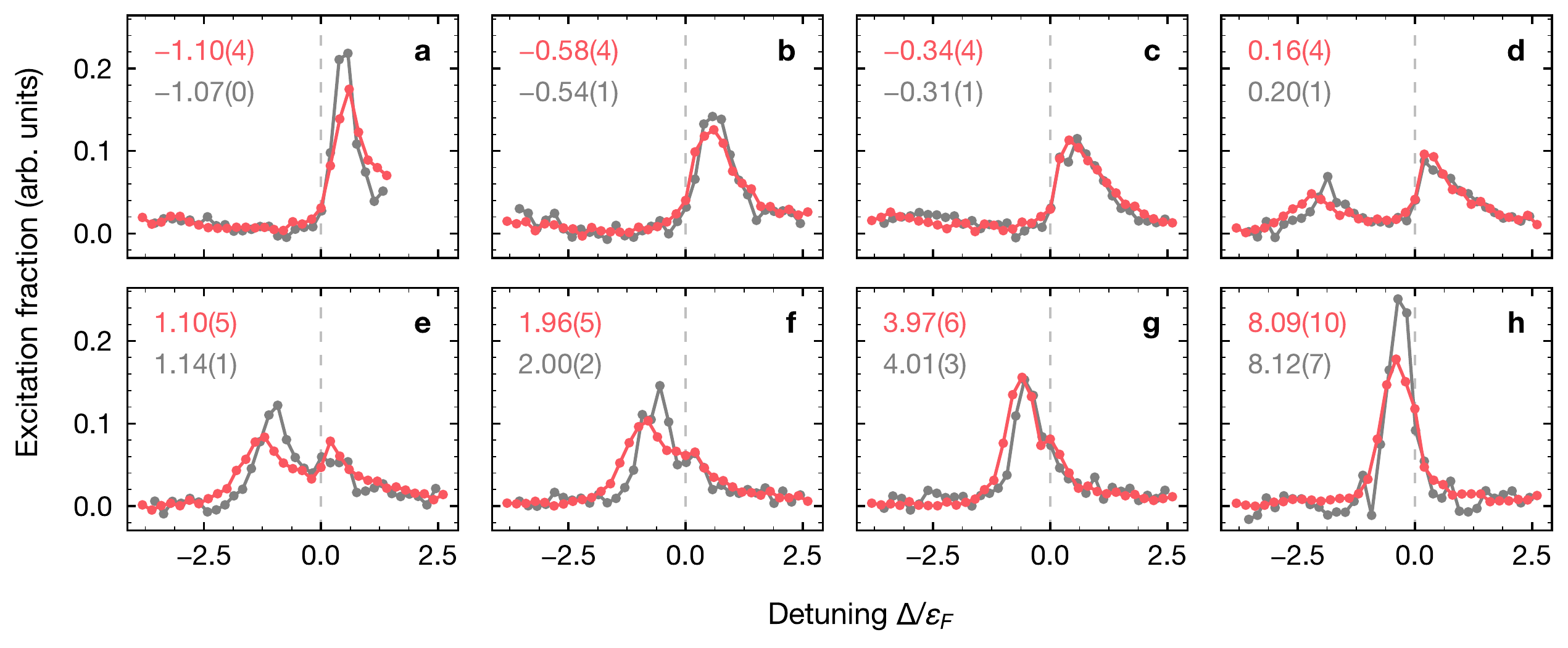}
  \caption{\label{fig:spectrum_su3_vs_su2}%
    Clock spectra in the presence (red circles and lines) and absence (gray circles and lines) of an additional Fermi sea in the $m_F=-5/2$ ground state at magnetic fields (a) $15\Gauss$, (b) $45\Gauss$, (c) $60\Gauss$, (d) $90\Gauss$, (e) $135\Gauss$, (f) $165\Gauss$, (g) $210\Gauss$, and (h) $255\Gauss$.
    Signal amplitudes are normalized to the integral of the full spectrum at the given magnetic field.
    Each data point is the average of two or three individual measurements and error bars are not shown to reduce visual clutter.
  The interaction parameters $\ln\left(\kappa_F a_\mathrm{2D}\right)$ are shown in the top left corner of each plot.
    Note that they differ at the same magnetic field due to slightly different effective Fermi energies $\epsilon_F = h \times 2.99(19)\kHz$ and $h \times 3.20(21)\kHz$.
    The minority fractions are $0.28(5)$ and $0.23(3)$, temperatures of the $\gup$ Fermi sea are $k_B T / \epsilon_F = 0.23(4)$ and $0.20(4)$.%
  }%
\end{figure}
For probing the effect of an additional Fermi sea in $\gdown$, we follow the same state preparation as described in Section~\ref{sec:state-preperation}.
The only difference is the optical pumping sequence which we adjust to yield a nearly balanced spin mixture of $m_F=\pm 5/2$.
Minority atoms are still prepared in $m_F=-3/2$.
The two Fermi seas interact repulsively which modifies the local density compared to the case without an additional Fermi sea.
We use the local density approximation and an expansion of the chemical potential for weak repulsive interactions to find the local Fermi energy $E_F(x, z)$. In our system, the Thomas-Fermi equation is given by
\begin{align}
  \mu_0 = \mu\left[n(x, z)\right] + \frac{m}{2} \left( \omega_x^2 x^2 + \omega_z^2 z^2 \right),%
  \label{eq:thomas-fermi-equation}
\end{align}
where $n(x, z)$ is the local atom density, $\mu(n)$ is the local chemical potential, and $\mu_0$ is fixed by $\int n(x, z) = N$, where $N$ is the total number of atoms.
We use the expansion of the chemical potential for weak repulsive interactions in 2D up to second order from Ref.~\cite{engelbrecht:1992},
\begin{align}
  \mu(g) = E_F \left[1 + 2g + 4g^2(1-\ln2) +\mathcal{O}\left(g^3\right) \right].%
  \label{eq:chem-pot-rep-int}
\end{align}
Here, $E_F$ is the local Fermi energy and $g$ is a dimensionless interaction parameter which is related to $\ln(\kappa_F a_\mathrm{2D})$.
In 2D, we use the following form to calculate the finite temperature chemical potential~\cite{kittel:2004}
\begin{align}
  \mu(T) = k_B T \ln \left(e^{T_F/T} - 1\right),%
  \label{eq:chem-pot-fin-temp}
\end{align}
where $T_F = E_F/k_B$ is the local Fermi temperature.
Numerically solving Eqs.~\eqref{eq:thomas-fermi-equation},~\eqref{eq:chem-pot-rep-int}, and \eqref{eq:chem-pot-fin-temp} yields the local density $n(x, z)$ and effective Fermi energy $\epsilon_F$ in the presence of interactions and finite temperature.
For the experimental parameters, we find an effective Fermi energy $\epsilon_F = h \times 2.99(19)\kHz$ compared to $h\times 3.26(20)\kHz$ in the absence of an additional Fermi sea.
The effective Fermi energy of the additional Fermi sea is given by $\epsilon_F^\downarrow = h\times 2.73(19)\kHz$.
For the temperature of the $\gdown$ Fermi sea, we find $k_B T/\epsilon_F = 0.23(4)$.

In the clock spectroscopy measurement shown in Fig.~\ref{fig:spectrum_su3_vs_su2}, we follow the same methods described in the main text and Section~\ref{sec:clock-spectroscopy} except for the modified state preparation.
We can still clearly identify the repulsive and attractive polaron branches but do not find any striking features in the spectrum compared to the configuration without the additional Fermi sea.
However, at intermediate magnetic fields we find a small shift of the attractive polaron energy towards smaller energies (see Fig.~\ref{fig:spectrum_su3_vs_su2}).
This shift towards smaller energies is in contrast to the results of our theoretical model (see Section~\ref{sec:dual-fermi-sea-conf}).
We point out that the energy shift is on the same level as our experimental resolution as well as the correction due to the repulsive interactions of the $\gdown{}$ and $\gup$ Fermi sea discussed above.
%
\begin{figure}
  \centering
	\includegraphics[width=\textwidth]{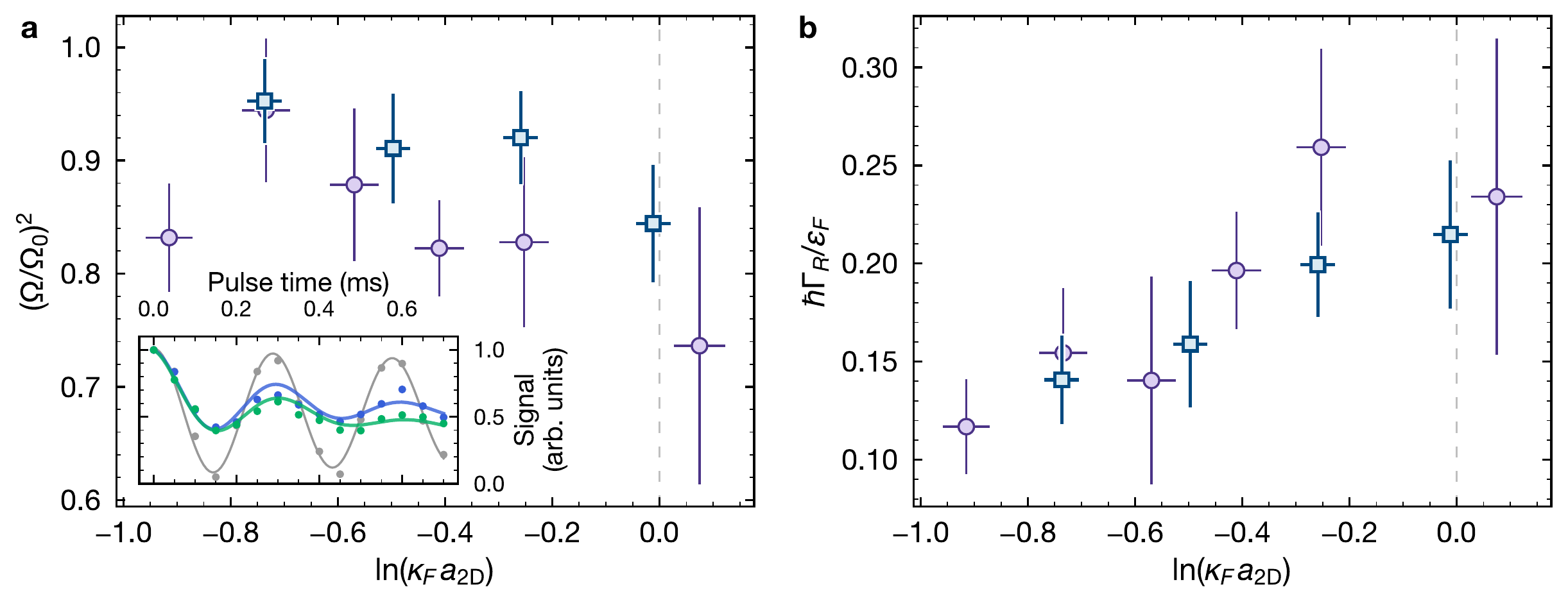}
  \caption{\label{fig:rabi_flops_su3_vs_su2}%
		(a) Quasiparticle residue $Z=(\Omega/\Omega_0)^2$ in the presence (blue squares) and absence [purple circles, see Fig.~3(a) of the main text] of an additional Fermi sea in the $\gdown$ state.
		The inset shows sample Rabi oscillations in the presence of an additional Fermi sea for interaction parameters $\ln\left(\kappa_F a_\mathrm{2D} \right) = -0.50(3)$ (blue points) and $-0.01(3)$ (green points), where solid lines are fits to the data points.
    The gray points and solid line correspond to a reference measurement without any background Fermi sea.
    (b) Damping $\Gamma_R$ of Rabi coupling as described by the fit function in Eq.~\eqref{eq:rabi-oscillations-fit-function}.
    Error bars indicate the fit error in $\Gamma_R$ and the uncertainty in $\ln\left(\kappa_F a_\mathrm{2D} \right)$.%
  }%
\end{figure}%

For the Rabi oscillations, we can only probe the repulsive polaron due to the finite tuning range of our laser used for the ``push'' pulses on the intercombination line.
For the repulsive polaron, the normalized Rabi frequency ${(\Omega/\Omega_0)}^2$ is slightly larger in the presence of an additional Fermi sea which could indicate a larger quasiparticle residue in this case.
The decoherence of Rabi oscillations $\Gamma_R$ is comparable for both configurations as shown in Fig.~\ref{fig:rabi_flops_su3_vs_su2}.

\section{Theoretical description}

\subsection{Two-body problem in a quasi-two-dimensional geometry}
In our theoretical modeling, we assume that the $^{173}$Yb atoms move in a uniform two-dimensional plane, while in the direction transverse to the plane we assume a harmonic potential $V(y)=\frac12m\omega_y^2y^2$.
We define the associated oscillator length $l_y=1/\sqrt{m\omega_y}$ (here and throughout this section we work in units where $\hbar=1$).
In this geometry, the center-of-mass and relative motion decouple.
Therefore, to describe the two atoms we consider the Hamiltonian of the relative motion, $\hat{\mathcal{H}} = \hat{H_0} + \hat{V}$.
Here, the non-interacting part is
\begin{align}
\hat H_0 = \sum_{\vect{k},n}2 \epsilon_{\vect{k}n} \ket{o,\vect{k}n}\bra{o, \vect{k}n} +  \sum_{\vect{k},n} (2\epsilon_{\vect{k}n} + \delta) \ket{c,\vect{k}n}\bra{c, \vect{k}n},
\end{align}
where $ \k$ is the in-plane relative momentum, $n$ denotes the transverse harmonic oscillator quantum number for the relative motion, and $2\epsilon_{\k n}=\k^2/m+n\omega_y$ is the non-interacting energy (measured from the zero-point energy).
The $^{173}$Yb system features two electronic orbitals ($\ket{g}$ and $\ket{e}$) as well as two nuclear spin states ($\ket{\downarrow}$ and $\ket{\uparrow}$).
In principle, there are four different configurations for two atoms, but for the situation of interest we only have two: The open channel $\ket{o}\equiv\ket{g\uparrow, e\downarrow}$ and the closed channel $\ket{c}\equiv\ket{e\uparrow, g\downarrow}$.
Here, $\delta$ is the detuning of the closed channel,
\begin{align}
    \delta\equiv \Delta \mu\, B
\end{align}
with $\Delta \mu$ given by Eq.~\eqref{eq:diff-zs-shift} and $B$ the magnetic field.
The interaction does not preserve the orbital configuration, but rather proceeds via the triplet ($+$) and singlet ($-$) channels, with $\ket{\pm}\equiv\frac1{\sqrt{2}}\left(\ket{o}\pm\ket{c}\right)$. Thus, the interaction part of the Hamiltonian takes the form
\begin{align}
\label{eq:hamint}
\hat V = \sum_{\k,\k',n,n'} \phi_n\phi_{n'}\left\{U_+ \ket{+,\k n}\bra{+, \k'n'}+U_- \ket{-,\k n}\bra{-, \k'n'}\right\},
\end{align}
where we assume contact interactions of strength $U_\pm$. Here,
\begin{align}
\phi_n=&\left\{\begin{array}{cl}
(-1)^{n/2} \frac{1}{(2\pi l_y^2)^{1/4}}\frac{\sqrt{n!}}{2^{n/2} (n/2)!},\quad& n \mbox{ even}\\
0, & n \mbox{ odd}
\end{array}
\right.
\end{align}
is the wave function of the transverse relative motion at zero separation.

The renormalization of the contact interaction in a quasi-two-dimensional geometry was carried out in Refs.~\cite{petrov:2001,bloch:2008}.
Here, we mainly follow the notation outlined in the review~\cite{levinsen:2015}, properly generalized to account for the orbital structure of the interactions.
We start by writing down the Lippmann-Schwinger equation for the two-body $T$ matrix
\begin{align} \label{eq:LSmat2}
\bra{\vect{k}'n'} \hat{T}(E)  \ket{\vect{k}n} = \bra{\vect{k}'n'} \hat{V} \ket{\vect{k}n} + \sum_{\vect{k}''n''} \bra{\vect{k}'n'} \hat{V} \ket{\vect{k}''n''} \bra{\vect{k}''n''} \frac{1}{E-\hat H_0+i0} \ket{\vect{k}''n''} \bra{\vect{k}''n''} \hat{T}(E) \ket{\vect{k}n},
\end{align}
where $E$ is the energy, and the infinitesimal positive imaginary part $+i0$ ensures that we consider the outgoing scattered part of the wave function.
Note that we suppress the indices related to the orbital configuration since we will consider this in a matrix representation.
Since the interaction part of the Hamiltonian, Eq.~\eqref{eq:hamint}, is independent of momentum, Eq.~\eqref{eq:LSmat2} factorizes as a function of momentum.
Thus, in the triplet-singlet basis we have
\begin{align}
\bra{\vect{k}'n'} \hat{V} \ket{\vect{k}n} =
\phi_n \phi_{n'}
\begin{pmatrix}
U_+ & 0 \\
0 & U_-
\end{pmatrix}
\equiv \phi_n \phi_{n'} \mathbf{V}.
\end{align}
On the other hand, the polarization bubble is most straightforward to evaluate in the open-closed channel basis where it takes the form
\begin{align}
    \sum_{\vect{k},n}  \bra{\vect{k}n} \frac{1}{E-\hat H_0+i0} \ket{\vect{k}n}=\sum_{\k,n}\begin{pmatrix}
\frac{\phi_n^2}{E-2\epsilon_{\k n}+i0} & 0 \\
0 & \frac{\phi_n^2}{E - 2\epsilon_{\k n}-\delta+i0}
\end{pmatrix}
\equiv\mathbf{\Pi}_{\rm{q2D}}(E).
\end{align}
With these definitions, it is straightforward to formally obtain the vacuum scattering $T$ matrix, which we write in the open-closed basis
\begin{align}
\mathbf{T}^{(\rm{vac})}(n_1,n_2;E)\equiv \bra{\vect{k_1}n_1} \hat{T}(E)  \ket{\vect{k_2}n_2}=\phi_{n_1}\phi_{n_2}\left[ \mathbf{R}\mathbf{V}^{-1}\mathbf{R} - \boldsymbol{\Pi}_{\rm{q2D}}(E) \right]^{-1}
\end{align}
for arbitrary relative momenta $\k_{1,2}$. Here,
\begin{align}
\mathbf{R}&=\frac1{\sqrt{2}}
\begin{pmatrix}
1 & 1 \\
1 & -1
\end{pmatrix}
\end{align}
is the involutory matrix that transforms between the triplet-singlet basis and the open-closed channel basis.

Following the steps in Ref.~\cite{levinsen:2015} (see also Refs.~\cite{petrov:2001,bloch:2008}) to evaluate the polarization bubble, we obtain
\begin{align}
    \mathbf{T}^{(\rm{vac})}(0,0;E)=\frac{2\sqrt{2\pi}}m\left[\mathbf{R}\begin{pmatrix} \frac{l_y}{a_+} & 0\\ 0 & \frac{l_y}{a_-}\end{pmatrix}\mathbf{R}-
    \begin{pmatrix}
    \mathcal{F}(-E/\omega_y) & 0 \\ 0 & \mathcal{F}((-E+\delta)/\omega_y)
    \end{pmatrix}
    \right]^{-1}
    \label{eq:Tq2D}
\end{align}
where we specialize to scattering between the lowest transverse levels, and consider excited states only in the intermediate virtual processes.
This is a valid assumption since in our experiment the transverse confinement frequency greatly exceeds all other relevant energy scales.
The function $\mathcal{F}$ takes the form~\cite{bloch:2008}
\begin{align}
    {\cal F}(x)=\int_0^\infty
		\frac{du}{\sqrt{4\pi u^3}}\left[1-\frac{e^{-xu}}{\sqrt{(1-e^{-2u})/2u}}\right].
\end{align}
The quantities $a_+$ and $a_-$ are the triplet and singlet scattering lengths, respectively.
While the contact interaction model, Eq.~\eqref{eq:hamint}, does not contain effective range corrections to the 3D scattering phase shifts, we could in principle have used a two-channel model for each of the singlet and triplet interactions to include this.
This procedure is straightforward, but cumbersome to write down.
The end result is, however, simple.
Indeed, we just have to make the replacement
\begin{align}
	a_\pm^{-1} \to a_\pm^{-1} - \frac{1}{2} m r_\pm \left(E- \frac{\delta}{2} + \frac{1}{2} \omega_y\right),
\end{align}
with $r_\pm$ the singlet and triplet effective ranges.
In practice, all theory curves are calculated by making this replacement and using the scattering lengths and effective ranges from Ref.~\cite{hoefer:2015}.

\begin{figure}[t!]
	\includegraphics[width=0.5\columnwidth]{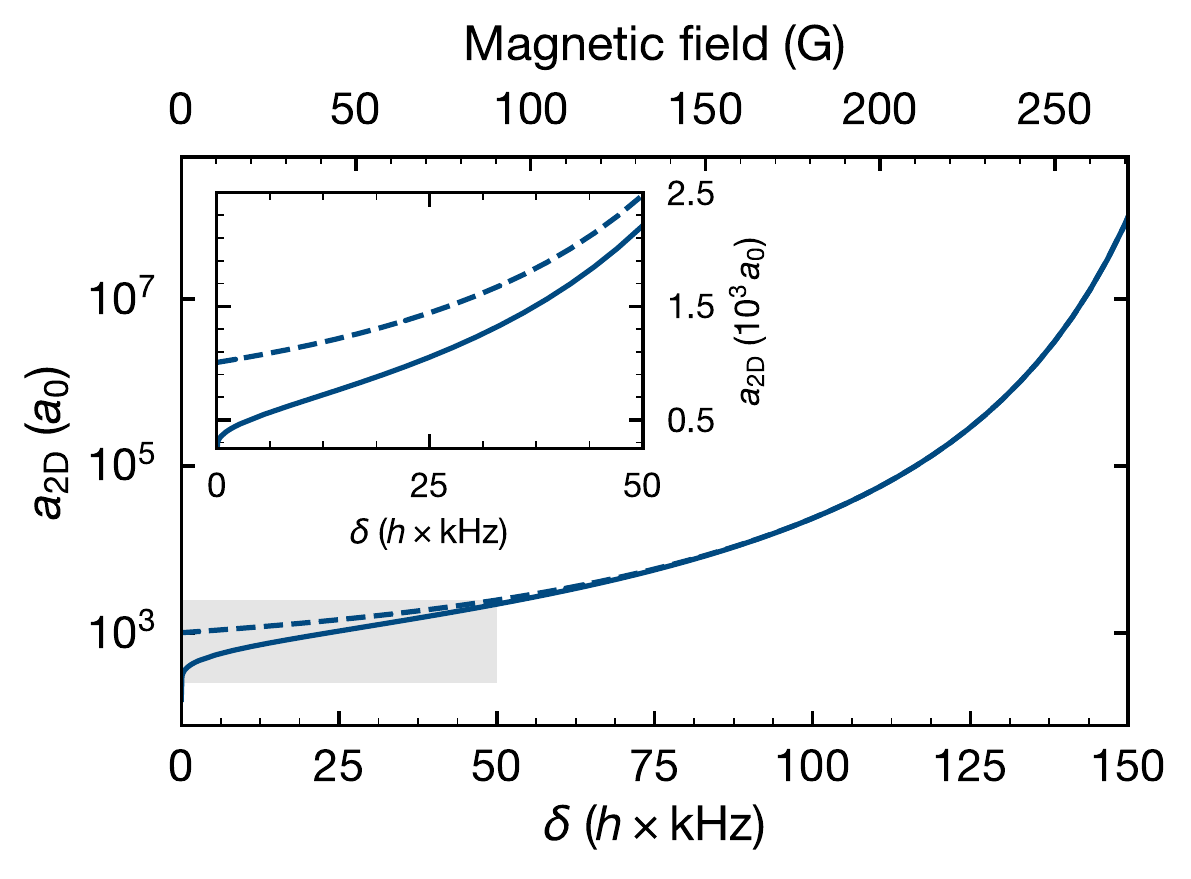}
  \caption{\label{fig:2d-scattering-length}%
    The 2D scattering length $a_\mathrm{2D}$ as a function of the closed channel detuning $\delta$ (magnetic field) for our typical experimental conditions with $l_y \simeq 750 a_0$.
		The solid line shows $a_\mathrm{2D}$ defined from the low-energy scattering amplitude [see Eq.~\eqref{eq:2d-scattering-length}], while the dashed line shows the $a_\mathrm{2D}$ that would be obtained by using the simple approximation $a_\mathrm{2D}=1/\sqrt{m \eb}$ where $\eb$ is the quasi-2D dimer binding energy defined in Eq.~\eqref{eq:eb}.
		For large $\delta$ (small $\eb$), the two results agree while for small $\delta$ (gray shaded area) there are rather large deviations as the inset shows.
  }
\end{figure}

Finally, we extract the two-body binding energy and the effective 2D scattering length from the $T$ matrix.
The two-body binding energy $\eb>0$ is found from the pole of the $T$ matrix, i.e., from the condition $\det[\mathbf{T}^{(\rm{vac})}(0,0;\eb)]^{-1} = 0$. Thus, $\eb$ satisfies
\begin{align}
	\left[\frac{2l_y}{a_+}-\mathcal{F}\left(\frac{\eb}{\omega_y}\right)-\mathcal{F}\left(\frac{\eb+\delta}{\omega_y}\right)\right]
	\left[\frac{2l_y}{a_-}-\mathcal{F}\left(\frac{\eb}{\omega_y}\right)-\mathcal{F}\left(\frac{\eb+\delta}{\omega_y}\right)\right]-
	\left[\mathcal{F}\left(\frac{\eb}{\omega_y}\right)-\mathcal{F}\left(\frac{\eb+\delta}{\omega_y}\right)\right]^2= 0.\label{eq:eb}
\end{align}
To extract an effective low-energy open-channel 2D scattering length, we compare the open-channel element of Eq.~\eqref{eq:Tq2D} with the expression for the low-energy scattering:
\begin{align}
    f_{\rm q2D}(E)\equiv m\mathbf{T}^{(\rm{vac})}_{11}(0,0;E)\simeq \frac{4\pi}{-\ln(ma_\mathrm{2D}^2E)+i\pi},
\end{align}
where the subscript on $\mathbf{T}$ indicate the corresponding matrix element.
Assuming that $\delta\gg |E|$ (i.e., that we are not close to $B=0$) we find
\begin{align}
  a_\mathrm{2D}=l_y\sqrt{\frac{\pi}{D}}\exp\left[-\sqrt{2\pi}\frac{l_y^2/(a_-a_+)-\frac12(l_y/a_-+l_y/a_+){\cal F}(\delta/\omega_y)}{l_y/a_-+l_y/a_+-2{\cal F}(\delta/\omega_y)}\right].\label{eq:2d-scattering-length}
\end{align}
Again, we can replace $a^{-1}_\pm\to a^{-1}_\pm-\frac14r_\pm m(\omega_y-\delta)$ to include the effective ranges in the triplet-singlet channels.
Fig.~\ref{fig:2d-scattering-length} shows the functional dependence of $a_\mathrm{2D}$ for our typical experimental parameters.
Here, we also see that the universal formula in the 2D limit, $\eb = 1/(m a_\mathrm{2D}^2)$, breaks down at small $\delta$ (low magnetic field) where the dimer binding energy becomes comparable to the transverse confinement.

\subsection{Impurity Green's function}
To investigate the many-body dressing of an impurity, we evaluate its Green's function which, in general, also takes the form of a matrix.
However, we will work under the assumption that we create at most a single excitation of the fermionic medium.
Such an approximation has proven extremely accurate since it was first introduced in Ref.~\cite{chevy:2006,combescot:2007}.
Within this approximation, the open and closed channel parts of the impurity Green's function completely decouple, and since our clock laser pulse injects the impurity into the open channel, we restrict ourselves to considering this channel.
Furthermore, since the transverse confinement frequency greatly exceeds both the Fermi energies and temperature, we assume that the impurity is only virtually scattered into excited bands of the transverse confinement such that we can consider an effective 2D Green's function.
We note that higher bands of the transverse confinement can in principle be taken into account using the formalism developed in \cite{levinsen:2012}.

Within our approximations, the open-channel retarded impurity Green's function satisfies the Dyson equation
\begin{align}
  G(\k, E)&=\frac1{E-\ek-\Sigma(\k,E)},
  \label{eq:green}
\end{align}
where we take the impurity to be in the lowest harmonic oscillator state, and we use the 2D single-particle energy $\ek\equiv \epsilon_{\k 0}$.
The open-channel self energy is
\begin{align}
  \Sigma(\k, E)=\sum_{\mathbf q}n_o(q)\mathbf{T}_{11}(\k+\q, E+\eq),
\end{align}
with $n_{o}(q) = \left\{1 + e^{\beta \left[q^2/(2m) - \mu\right]}\right\}^{-1}$ the finite-temperature Fermi-Dirac distribution of the background Fermi sea in $\ket{g,\uparrow}$ and $\mathbf{T}$ the in-medium $T$ matrix.
The latter is related to the vacuum $T$ matrix, Eq.~\eqref{eq:Tq2D}, via
\begin{align}
  \mathbf{T}^{-1}(\k,E)=\left[\mathbf{T}^{\rm (vac)}(E-\ek/2)\right]^{-1}
  +\left(
  \begin{array}{cc}
  \Delta\Pi_o(\k,E)
  & 0 \\
  0 & \Delta\Pi_c(\k,E-\delta)
  \end{array}
  \right).
\end{align}
Note that we have dropped the transverse harmonic oscillator indices on the vacuum $T$ matrix since these have been set to 0. The functions $\Delta\Pi_{o,c}$ are the differences between the medium and vacuum $T$ matrices:
\begin{align}
  \Delta\Pi_{o,c}(\k,E) & = \sum_\q \frac{n_{o,c}(q)}{E+\ek-\eq-\epsilon_{\k-\q}+i0},
\end{align}
where $n_c(q)$ is the Fermi-Dirac distribution of the additional Fermi sea in state $\ket{g,\downarrow}$ which must be taken into account for the configuration described in Sec.~\ref{sec:twoFS}.
We extract all quasiparticle properties from the impurity Green's function.

\subsection{Polaron spectral response}
The spectral response measured in experiment corresponds to the impurity spectral function $A(E)$, which we calculate from the Green's function in Eq.~\eqref{eq:green} as
\begin{align}
  A(E) & = -\frac1\pi\mbox{Im}[G(\0,E)],
\label{eq:Aomega}
\end{align}
where we neglect the initial distribution of impurity momenta.
We extract the attractive and repulsive polaron energies $E_-$ and $E_+$, respectively, from the peak values of the spectral function.
The results are shown in Figs.~1(a) and 2(a) of the main text.

\begin{figure}[t!]
	\includegraphics[width=\columnwidth]{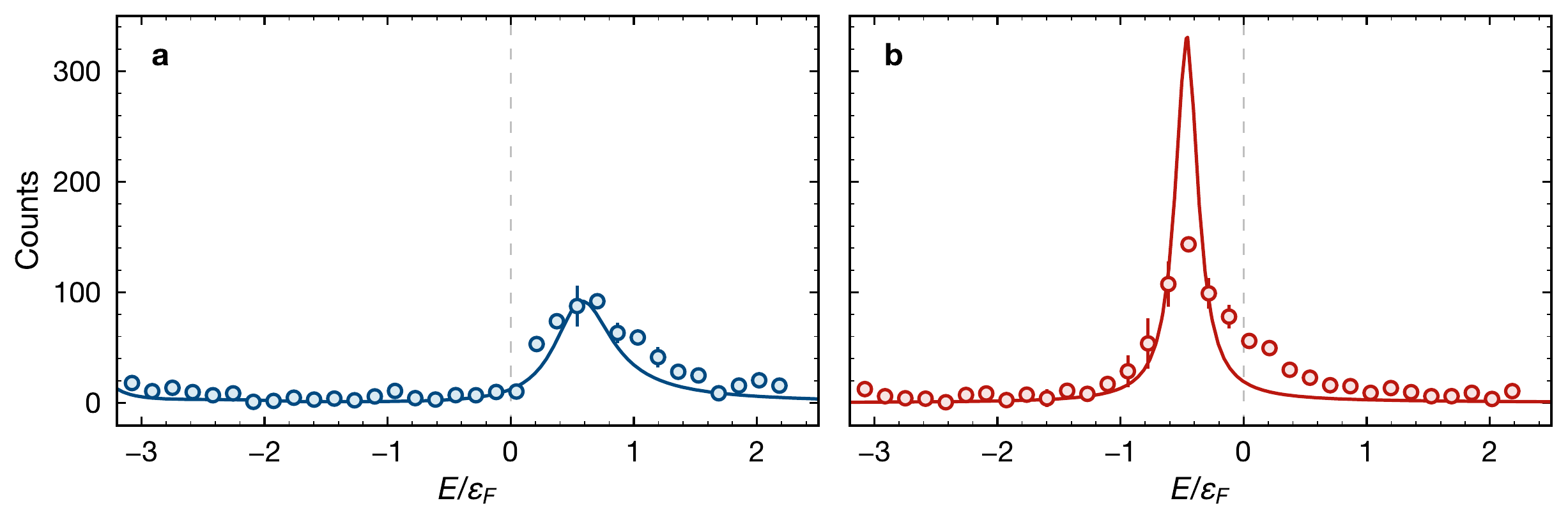}
  \caption{\label{fig:spectra}%
    Spectral response at fixed magnetic fields (a) $B=45$G [repulsive side with $\ln(\kappa_F a_\mathrm{2D}) = -0.38(2)$] and (b) $B=210$G [attractive side with $\ln(\kappa_F a_\mathrm{2D}) = 4.17(5)$].
		The experimental data (filled circles) is compared with the theoretical predictions (solid lines) obtained from Eq.~\eqref{eq:Aomega} at $T/T_F=0.17$ and with a small Lorentzian broadening of $0.11\epsilon_F$, which corresponds to the Fourier limit of the excitation pulse.
    Note that the overall amplitude of the theoretical spectrum is scaled to the height of the repulsive polaron peak in the experiment.
    To better compare the shapes of the peaks, we have shifted the theoretical curves down in energy by $0.2\epsilon_F$.
  }
\end{figure}

Fig.~\ref{fig:spectra} shows sample spectra recorded at fixed magnetic fields of (a) 45G and (b) 210G compared with the results of our theory.
We see that while the width of the repulsive polaron is approximately captured within our model, the attractive polaron peak is somewhat broader in experiment than in the theoretical model.

\subsection{Quasiparticle residue}
From the Rabi oscillations observed in experiment, we extract the quasiparticle residue, $Z$, defined as the squared overlap between the non-interacting state of the impurity plus medium with that of the strongly interacting system.
Similar to the scenario in alkali-atom experiments~\cite{kohstall:2012,scazza:2017}, this overlap in reality compares states involving the impurity atom in two different spin states.
However, the Rabi coupling provides the spin-flip operator that is necessary to simply relate the overlap to that of interacting and non-interacting states, without making explicit reference to the spin state.
In contrast to the case of alkali atoms, the interacting state of the impurity atoms involves two orbitals, but since the overlap with the non-interacting state only involves the open channel it can be extracted from the open-channel Green's function.

Specifically, the Rabi frequency $\Omega$ is reduced by a factor $\sqrt{Z}$~\cite{kohstall:2012} from its bare value $\Omega_0$ due to the dressing of the impurity by excitations of the medium.
We calculate the residue using the standard expression
\begin{align}
  Z_\pm^{-1}=1-{\rm Re}\left[\left.\frac{\partial\Sigma(\0,E)}{\partial E}\right|_{E=E_\pm}\right].
\end{align}
The results are shown in Fig.~3(a) of the main text, where we see that the measured repulsive (attractive) polaron residues are systematically above (below) the theoretical prediction.

\begin{figure}[t!]
	\includegraphics[width=0.5\columnwidth]{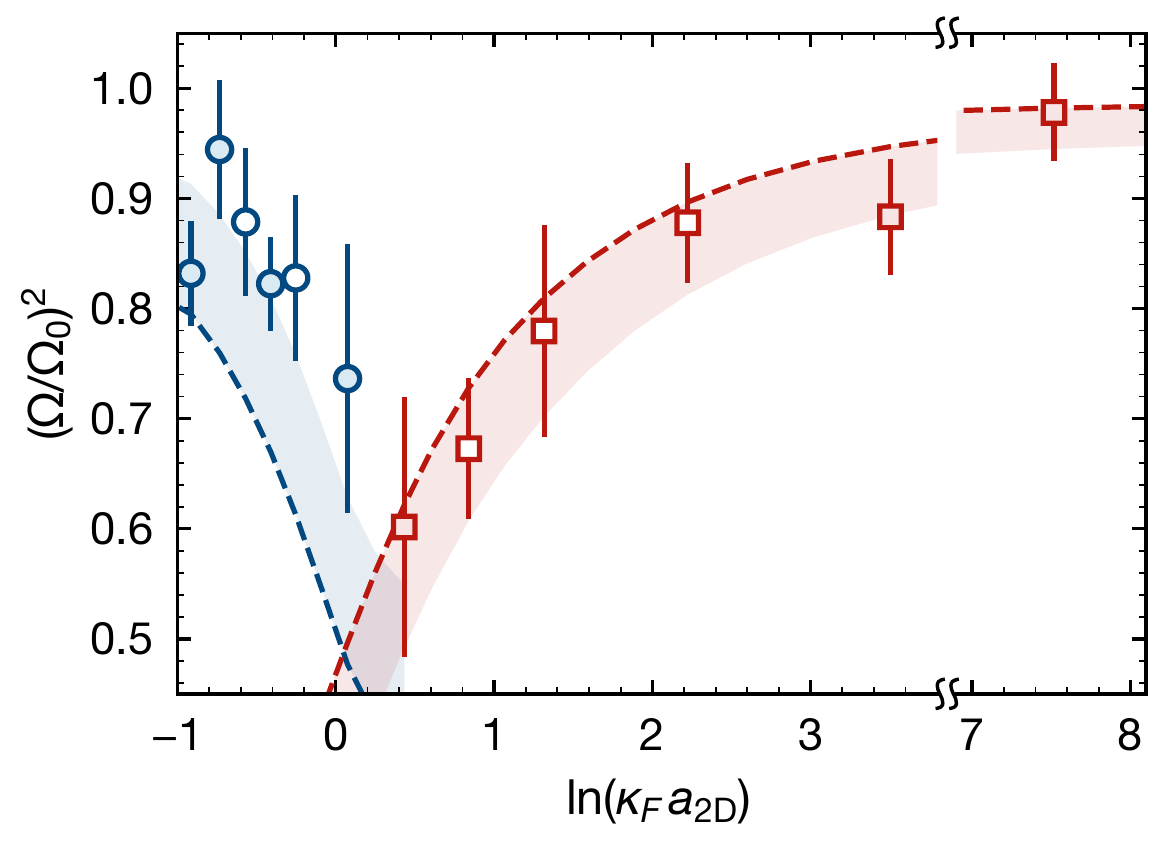}
  \caption{\label{fig:residue-bounds}%
    The quasiparticle residue extracted from fits to Rabi oscillations as in Fig.~3(a) of the main text. Blue circles (red squares) correspond to the repulsive (attractive) polaron.
    The residues from our theory without considering initial state interactions are shown as dashed lines, while the shaded bands illustrate the range of possible values that $Z$ can take once these are taken into account.
  }
\end{figure}

The main source of discrepancy between theory and experiment for the residue shown in Fig.~3(a) is likely to be the repulsive interactions between the initial $\ket{g,0}$ state and the Fermi sea in state $\ket{g,\up}$.
Such repulsive interactions will respectively enhance and reduce the overlap of the initial state with the repulsive and attractive polarons in the final state.
We can estimate the maximum amount this overlap can change as follows.
Assume that a polaron $\ket{\Psi_p}$ with residue $Z_p$ has a wave function consisting of only two terms, the state that overlaps with the non-interacting ground state $\ket{0}$ and an incoherent background $\ket{\rm inc}$:
\begin{align}
  \ket{\Psi_p} = \sqrt{Z_p}\ket{0}+e^{i\varphi_p}\sqrt{1-Z_p}\ket{\rm inc},
\end{align}
where we assume that both $\ket{0}$ and $\ket{\rm inc}$ are normalized and $e^{i\varphi_p}$ is a phase. Within this approximation, we would thus arrive at the maximum (minimum) overlap between the repulsive (attractive) polaron states $\ket{\Psi_+}$ ($\ket{\Psi_-}$) and the initial (repulsive polaron) state $\ket{\Psi_i}$:
\begin{align}
  |\bra{\Psi_i}\!\Psi_+\rangle| & \lesssim  \sqrt{Z_+}+\sqrt{1-Z_i}\sqrt{1-Z_+},
  \nn \\
  |\bra{\Psi_i}\!\Psi_-\rangle| & \gtrsim  \sqrt{Z_-}-\sqrt{1-Z_i}\sqrt{1-Z_-},
\end{align}
with the initial-state residue $Z_i\simeq0.98$. We show these estimates in Fig.~\ref{fig:residue-bounds}, and we see that this can improve the agreement between theory and experiment substantially.
A similar effect was found in Ref.~\cite{scazza:2017} for $^6$Li atoms.

\subsection{Damping of Rabi oscillations}
Another observable is the width $\Gamma_\pm$ of each polaron peak in the spectral function, which can be computed from the self energy for the attractive and repulsive branches:
\begin{align} \label{eq:width}
  \Gamma_\pm = -Z_\pm\rm{Im}[\Sigma(\0,E_\pm)].
\end{align}
The result $\Gamma_+$ for the repulsive polaron is plotted in Fig.~3(b) of the main text and is seen to match the damping rate of the Rabi oscillations observed in experiment. However, for the attractive polaron, the damping rate is much larger than that predicted from Eq.~\eqref{eq:width}.
This enhanced damping could be because the Rabi frequency in experiment is comparable to or larger than the Fermi energy, which has been theoretically shown to affect the Rabi oscillations of the attractive polaron~\cite{parish:2016}.

To understand the damping rate of the attractive polaron, we consider a simple three-state model of the Rabi oscillations that involves the non-interacting initial state $\ket{\Psi_i}$, the final attractive polaron state $\ket{\Psi_-}$, and a state $\ket{\Psi_{\rm con}}$ that represents the continuum of states in the final spectrum that are orthogonal to $\ket{\Psi_-}$.
Taking the clock laser to be resonant with the attractive polaron, the Hamiltonian for the three-state system is, in matrix form,
\begin{align}
  H=
  \begin{pmatrix}
  0 & \sqrt{Z} \frac{\Omega_0}{2} & \sqrt{1-Z} \frac{\Omega_0}{2} \\
  \sqrt{Z} \frac{\Omega_0}{2} & 0 & 0 \\
  \sqrt{1-Z} \frac{\Omega_0}{2} & 0 &  -i \Gamma_{\rm con}
  \end{pmatrix} .
\end{align}
Here, we have assumed that the continuum is dominated by its spectral width $\Gamma_{\rm con} \sim \epsilon_F$ since it does not correspond to a well-defined peak in the spectrum. We have also required that the spectrum obeys the sum rule $|\langle \Psi_i |\Psi_-\rangle|^2 + |\langle \Psi_i | \Psi_{\rm con}\rangle|^2 = 1$, so that $|\langle \Psi_i | \Psi_{\rm con}\rangle| = \sqrt{1-Z}$ with $Z =|\langle \Psi_i |\Psi_-\rangle|^2$.

When the final state is non-interacting, then $Z = 1$ and we obtain the standard result for Rabi oscillations of the initial-state fraction $N_i$ as a function of time $t$,
\begin{align}
  N_i(t) = \frac{1}{2}\left[1+\cos(\Omega_0 t)\right] .
\end{align}
However, the continuum can significantly affect the oscillations when $Z < 1$ and $\Omega_0$ is comparable or larger than $\epsilon_F$, as is the case in experiment.
In particular, if we assume that $\Omega_0 \gg \Gamma_{\rm con}$ and $Z$ is close to 1, a perturbative calculation yields
\begin{align}
  N(t) \simeq \frac{e^{-\Gamma_R t}}{2}\left[1+\cos\left(\sqrt{Z}\Omega_0 t\right)\right] ,
\end{align}
where the Rabi damping rate $\Gamma_R \simeq (1-Z) \Gamma_{\rm con}$.
This is consistent with the results shown in Fig.~3(b) of the main text.
While we do not have a precise value for the width of the continuum $\Gamma_{\rm con}$, we expect it to be comparable to the Fermi energy $\epsilon_F$ of the medium.

\begin{figure}[t!]
	\includegraphics[width=0.5\columnwidth]{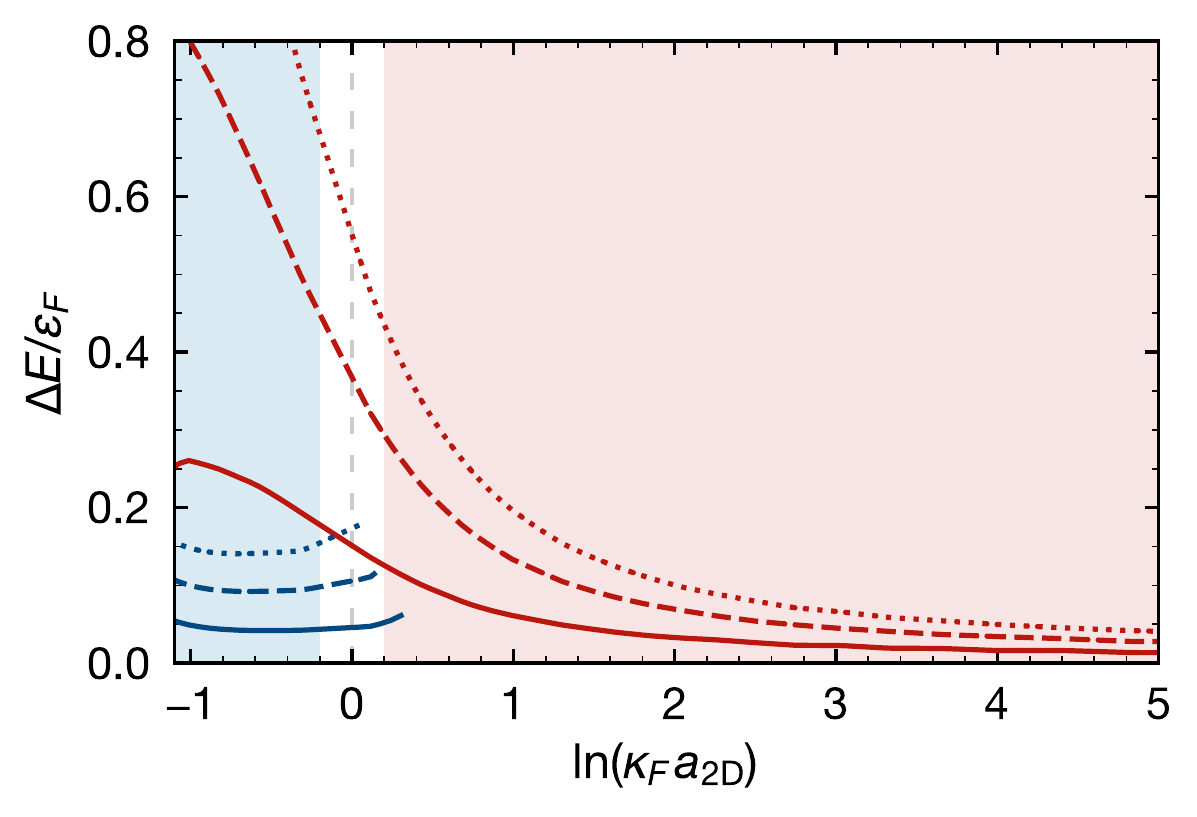}
  \caption{\label{fig:energy-shifts}%
    Theoretically predicted energy shifts of repulsive (blue lines) and attractive (red lines) polarons for $\epsilon_F = h \times 3.0\kHz$ and an additional Fermi sea in $\gdown{}$ with an effective Fermi energy $\epsilon_F^\downarrow = h \times 2.7\kHz$ (solid), $2 \epsilon_F^\downarrow$ (dashed), and $3 \epsilon_F^\downarrow$ (dotted).
    The blue (red) shaded area denotes the interaction parameters where we experimentally find a sufficient contrast of the repulsive (attractive) polaron peak.}
\end{figure}

\subsection{Dual Fermi sea configuration}\label{sec:dual-fermi-sea-conf}
Finally, we comment on the dual Fermi sea configuration discussed experimentally in Sec.~\ref{sec:twoFS}.
The motivation for this investigation is that this scenario corresponds to an effective frustration of interactions close to the orbital Feshbach resonance.
Indeed, while the impurity in state $\ket{g,0}$ does not directly (strongly) interact with the Fermi sea in state $\ket{g,\downarrow}$, the presence of a Fermi sea in the weakly-detuned closed channel provides an effective Pauli blocking of the open-channel interactions.
This is fundamentally different from the usual Feshbach resonances~\cite{chin:2010}, where the closed channel is detuned by an energy that far exceeds all scales relevant to the physics of interest.
Thus, our experiment provides the first steps towards realizing a ``frustrated'' Feshbach resonance.

Fig.~\ref{fig:energy-shifts} shows our predicted energy shifts in the ``frustrated'' configuration.
We see that we expect the energy of both polaron branches to increase.
For the experimental Fermi energy in the closed channel, the expected energy shifts (solid lines in Fig.~\ref{fig:energy-shifts}) are below $0.15\epsilon_F$ within the range of magnetic fields where we have sufficient contrast, which may be the reason why these shifts are not observed in experiment (see Fig.~\ref{fig:spectrum_su3_vs_su2}).
However, as also illustrated in Fig.~\ref{fig:energy-shifts}, we expect the energy shift to strongly increase with the $\ket{g,\downarrow}$ Fermi energy.
In particular, the increased energy of the repulsive branch could potentially stabilize a ferromagnetic phase~\cite{massignan:2014}.

\enlargethispage{\baselineskip}

\bibliography{references}